\documentclass[preprint,amsmath,amssymb,aps]{revtex4}
\usepackage{graphicx}
\usepackage{hyperref}
\usepackage{epstopdf}
\usepackage{slashed}
\usepackage{rotating}
\usepackage{float}
\usepackage{array}
\newcommand{\be}{\begin{equation}}
\newcommand{\ee}{\end{equation}}
\newcommand{\bea}{\begin{eqnarray}}
\newcommand{\eea}{\end{eqnarray}}
\newcommand{\ba}{\begin{array}}
\newcommand{\ea}{\end{array}}

\begin{document}

\title{Axial-vector charges of the spin $\frac{1}{2}^+$ and spin $\frac{3}{2}^+$ light and charmed  baryons in the SU(4) chiral quark constituent model}

\author{Harleen Dahiya}
\email{dahiyah@nitj.ac.in}
\affiliation{Department of Physics, Dr. B.R. Ambedkar National
	Institute of Technology, Jalandhar, 144027, India}
\author{Suneel Dutt}
\email{dutts@nitj.ac.in}
\affiliation{Department of Physics, Dr. B.R. Ambedkar National
	Institute of Technology, Jalandhar, 144027, India}
\author{Arvind Kumar}
\email{kumara@nitj.ac.in}
\affiliation{Department of Physics, Dr. B.R. Ambedkar National
	Institute of Technology, Jalandhar, 144027, India}
\author{Monika Randhawa}
\email{monika@pu.ac.in}
\affiliation{University Institute of Engineering and Technology,
	Panjab University, Chandigarh, 160014, India}

\date{today}
\begin{abstract}
	Following the first clear evidence of the presence of intrinsic charm contribution in the proton, the axial-vector charges of the light and charmed baryons are investigated in the framework of $SU(4)$ chiral constituent quark model after including the explicit contributions from the $u\bar u $, $d\bar d $, $s\bar s $ and $c\bar c $ fluctuations.    The axial-vector charges having physical significance correspond to the generators of the $SU(4)$ group with flavor singlet $\lambda^0$, flavor isovector $\lambda^3$, flavor hypercharge $\lambda^8$ and flavor charmed $\lambda^{15}$ combinations of axial-vector current at zero momentum transfer.  In contemplation to further understand the $Q^2$ dependence of these charges, we have used the conventionally established dipole form of parametrization. The baryons considered here are the spin $\frac{1}{2}^+$ and spin $\frac{3}{2}^+$  multiplets decomposed further depending on the charm content of baryons.


\end{abstract}

\maketitle

\section{Introduction}

Ever since the first direct indication for the point like constituents in the nucleon from the measurements of polarized structure functions in the deep inelastic scattering (DIS) experiments \cite{emc,smc,adams,hermes}, showing surprisingly that only 30\% of the proton spin is carried by the constituent quarks, a lot of experiments have been conducted to probe the structure of proton. Even though these experiments have provided sufficient evidence on the polarized structure functions of the nucleon, the decomposition of the proton's spin in terms effective degrees of freedom still persists to be a major unsettled and inconclusive issue in high energy spin physics. In pursuit to understand the spin structure of the nucleon, several experimental and theoretical studies have been stimulated and the spin decomposition of nucleon in terms of intrinsic spin and orbital angular momentum of  the quark sea and gluons is regarded as one of the vital areas in hadronic physics in the present day. This is in fact one of the primary goals of the upcoming Electron-Ion Collider (EIC) project.

Since the composite baryons are bound by increasingly strong forces for decreasing momentum transfer, different current probes are needed to provide a profound understanding in inferring the internal structure. The electromagnetic current probes the  Dirac and Pauli form factors which are well understood over a broad region of momentum transfer squared $Q^2$. On the other hand, the isovector axial-vector current furnishes information on the internal spin structure but the study has been rather limited. The first moment of the spin structure function is further related to the mixtures of the axial-vector coupling constants. The flavor singlet component $g^0$ is one of  the fundamental parameter to determine the spin structure of the nucleon and can be associated with the total quark spin content, the flavor isovector (triplet) component $g^3$ determines exactly the neutron $\beta-$decay parameters and has received much consideration in the past, the flavor hypercharge axial (octet) component $g^8$ delivers the information on the hyperon $\beta$ decay parameters.
The study of axial charges play a significant role in exploring non-perturbative hadronic physics as they are connected to the form factors which are used to study the internal structure of hadrons.
One can connect the axial charges to the order parameter of spontaneous breaking of chiral symmetry \cite{choi2010}.
The axial-vector charges  can be directly related to the polarized deep inelastic scattering data hence shedding light on the contribution of quarks to spin of baryons. The study of axial vector charges is important for both weak and strong interaction sectors. These axial-vector charges arising from the axial-vector current are basically mixtures of the spin polarizations, $\Delta u$, $\Delta d$ and $\Delta s$ and they  carry complete data on the spin structure of the baryons. They  can further be related to certain prominent and renowned sum rules such as Bjorken sum rule (BSR) \cite{bjorken} and Ellis-Jaffe sum rule (EJSR) \cite{ellis}.  

In the past, experiments measuring electromagnetic and weak form factors from the elastic scattering of electrons have already given indications for the presence of strange quarks in the nucleon. The results are available for SAMPLE at MIT-Bates \cite{sample}, G0 at JLab \cite{g0}, PVA4 at MAMI \cite{a4} and HAPPEX  at JLab \cite{happex}. Further, the violation of EJSR has established the crucial role of quark sea in  the nuclen spin structure. To take this understanding a step further, semi-inclusive DIS (SIDIS) experiments have been performed to measure the sea quark contributions explicitly and separate them from the constituent quark contributions. Most recently, the heavy quark contribution in the dynamical properties of hadrons has been discussed \cite{stan-cheavy} as quantum chromodynamics (QCD) predicts the existence of both intrinsic and extrinsic heavy quark contributions to the fundamental structure of hadrons. The NNPDF collaboration has recently established the existence of intrinsic charm in the proton from structure function measurements \cite{charm-2022}. In the context of axial-vector charges, our information on the heavy baryons is rather limited because of their short lifetimes and  it is difficult to measure their properties experimentally. Very recently, results have been reported for light and singly  heavy baryons in the chiral quark soliton model \cite{cqsm-Kim-2022}.

The conventional expectation would be to perform the calculations from first principles of  QCD. However, the problem of confinement restricts our knowledge and the study on axial-vector charges of the heavy quarks has been rather limited. Some lattice QCD calculations have been reported \cite{lattice-2016} on the hyperons and charmed baryons but there is lot of scope for refinements to be done  on the broader question of axial-vector charges and the challenges faced because of confinement. Important constraints are furnished by partial conservation of axial-vector current (PCAC) \cite{pcac-ref1,pcac-ref2,buchmann-axial-mass}.
One of the extremely promising nonperturbative approach which provides, among other insights, unique and important information on the distribution structure of the quarks in the baryons is the chiral constituent quark model ($\chi$CQM) \cite{manohar}. It is rested upon the elementary idea of chiral symmetry breaking which takes place at a much smaller distance scale as compared to that of the confinement scale. This symmetry breaking directs the massless quarks, coupled with internal Goldstone bosons (GBs), to attain dynamical mass  \cite{cheng,johan,song,hd}. This in result provides crucial indications to interpret  the nonperturbative aspects of QCD physically. Apart from giving a feasible solution to the static properties like ``proton spin crisis'' \cite{hd}, magnetic moments of light
octet and decuplet baryons \cite{hdmagnetic}, implicit Gottfried Sum Rule violation
\cite{hdasymmetry}, satisfying the Coleman-Glashow sum rule, hyperon $\beta$ decay parameters \cite{nsweak}, strangeness content in the nucleon \cite{hds}, magnetic moments of ${\frac{1}{2}}^-$ octet baryon resonances \cite{nres}, magnetic moments of ${\frac{1}{2}}^-$ and ${\frac{3}{2}}^-$ $\Lambda$ resonances \cite{torres}, charge radii \cite{charge-radii}, quadrupole moment \cite{quad}, etc.. This approach has been further extended to anticipate the important role played by the intrinsic charm content in the nucleon spin \cite{hdcharm} and to use this for the calculations of the magnetic moment and charge radii of spin ${\frac{1}{2}}^+$ and spin ${\frac{3}{2}}^+$ charm baryons including their radiative decays \cite{nscharm,chargeradii-charm}. All the results agree with the latest experimental data available \cite{PDG}. In perspective of the above progresses, extending the model to calculate the axial-vector charges of the light and charmed baryons becomes desirable.

The purpose of the present communication is to evaluate the axial-vector charges of the light and charmed baryons in the framework of $SU(4)$ $\chi$CQM. Results have been obtained for the axial-vector charges which have physical significance corresponding to the generators of the $SU(4)$ group with flavor singlet $\lambda^0$, flavor isovector $\lambda^3$, flavor hypercharge $\lambda^8$ and flavor charmed $\lambda^{15}$ combinations of axial-vector current at zero momentum transfer.
The axial-vector charges $g^0$, $g^3$, $g^8$  and $g^{15}$ can be expressed as mixtures of the spin polarizations at zero momentum transfer. They have been investigated for two sets of particles in the SU(4) representation. First, the case of  spin $\frac{1}{2}^+$ mixed symmetry 20-plet with the light baryon octet (8,0), singly charmed sextet (6,1) as well as anti-triplet ($\bar{3}$,2) and triplet (3,2) with doubly charmed baryons have been calculated. Second, the spin $\frac{3}{2}^+$  symmetric 20-plet with the light baryon decuplet (10,0), singly charmed sextet (6,1), doubly charmed triplet (3,2) and triply charmed singlet (1,3). In order to further understand the internal structure of baryons and the dynamics of quarks inside them by viewing the constituent quarks as spatially extended particles, the explicit contributions of the constituent quarks to the axial-vector charges can be calculated in the baryons. The $Q^2$ dependence of these charges has been explored for the range $0\leq Q^2\leq 1$ GeV$^2$ using the conventionally established dipole form of parametrization which will shed a refreshing new light on the issue of spin structure and confinement in baryons.

\section{Axial-Vector Charges for the spin $\frac{1}{2}^+$ and spin $\frac{3}{2}^+$  charmed baryon multiplets}

The axial-vector operator constituting the quark field for the spin $\frac{1}{2}^+$ and spin $\frac{3}{2}^+$  charmed baryon multiplets can be defined  as 
\be
A^{\mu,a}={\bf \overline{q} (x)}\gamma^\mu \gamma_5\frac{\lambda^a}{2} {\bf q(x)}, \ee
where ${\bf q(x)}$ designates the flavor space quark field for the light and charm quarks ${\bf q}=(u,d,s,c)$. Here $\lambda^a$ ($a=1,2,..15$) are the well known Gell-Mann matrices describing the flavor $SU(4)$ structure of the light and charm quarks. A unit matrix $\lambda^0(=\sqrt{\frac{2}{3}}I)$ can be conveniently introduced in addition to these
matrices giving the axial-vector charge corresponding to the flavor singlet current $g^{0}$. In the present context of axial-vector charges, we can construct only those matrices from the generators of $SU(4)$ gauge group which have diagonal representation. We have the generator $\lambda^3$  corresponding to the flavor isovector (triplet) current,  $\lambda^8$ corresponding to the flavor hypercharge axial-vector (octet) current and the other diagonal generator $\lambda^{15}$ which respectively give the axial-vector charges as $g^{3}$, $g^{8}$ and $g^{15} $\cite{pcac-ref1,pcac-ref2}. 

Using this operator, the matrix element of the axial-vector current can be parameterized in terms of the axial-vector charges.
The multiplet numerology for the subset of baryons
belonging to {\it SU}(4) flavor multiplets is given as
\be 4\times 4 \times 4= 20{_S}+20{_{M}} +20{_{M}}+\bar 4,
\ee where the mixed symmetry 20-plet consists
of 8+6+$\bar 3$+3 baryons flavor states.
The spin $\frac{1}{2}^+$ baryon multiplet includes the light baryon octet at the first level, singly charmed symmetric sextet and anti-symmetric anti-triplet at the second level. The top level consists of a triplet with doubly charmed baryons. The axial-vector current for this case is expressed as \cite{hdaxial,cqsm-Kim-2005}
\be \langle
B^{\frac{1}{2}^+}(p', J_z')|A^{\mu,a}|B^{\frac{1}{2}^+}(p,J_z)\rangle=\bar u(p', J_z') \left[
\gamma^\mu \gamma_5 G^a_A(Q^2)+ \frac{q^\mu}{2M_B} \gamma_5 G^a_P(Q^2) \right]  u(p, J_z)\,, \label{Amu12} \\
\ee where  $M_B$ is the baryon mass. The Dirac spinors of the initial (final) baryon states are presented as  $u(p)$ ($\bar u(p')$) and the four momenta transfer as $Q^2 = -q^2$, where $ q \equiv
p- p'$.  The functions $G^a_A(Q^2)$ and $G^a_P(Q^2)$ ($a=0,3,8,15$)
are the axial-vector and the induced pseudoscalar form factors respectively. 

 For the case of spin $\frac{3}{2}^+$ baryon multiplet we have the symmetry
 20-plet consisting of 10+6+3+1 with  the light baryon decuplet as the first level, singly charmed symmetric sextet at the second level, doubly charmed triplet at the third level and the triply charmed singlet at the top. The axial-vector current for this case is expressed as \cite{cqsm-Kim-2020}
\be \langle
B^{\frac{3}{2}^+}(p', J_z')|A^{\mu,a}|B^{\frac{3}{2}^+}(p,J_z)\rangle=\bar u_{\rho}(p',J_z') \left[
\gamma^\mu \gamma_5 G^a_{A}(Q^2) \eta^{\rho \sigma}+ \frac{q^\mu}{2M_B} \gamma_5 G^a_P(Q^2)\eta^{\rho \sigma} \right]  u_{\sigma}(p,J_z)\,, \label{Amu32} \\
\ee 
where $u^{\rho}(p,J_z)$ represents the  Rarita-Schwinger spinor describing the  spin $\frac{3}{2}^+$ baryon multiplet. It is a tensor product between a Dirac spinor and a first rank tensor. The metric tensor $\eta^{\rho \sigma}$ of Minkowski space is  expressed as $\eta^{\rho \sigma}$=diag$(-1,1,1,1)$.  
A spin$-1$ field or Lorentz vector can be constructed from the Dirac spinor $u(p)$ and the  combination of the polarization vector of the spin$-1$ field with the Dirac spinor of the spin $\frac{1}{2}$ field can characterize the Rarita-Schwinger spinor.
The Rarita-Schwinger spinors of the initial (final) spin $\frac{3}{2}^+$ baryon states are presented as $u^{\sigma}(p,J_z)$ ($\bar u_{\rho}(p',J_z')$). The induced pseudoscalar form factors $G^a_P(Q^2)$ from Eqs. (\ref{Amu12}) and (\ref{Amu32}) not being relevant have been ignored for the present work.

In general, one can extract the axial-vector charge corresponding to the axial-vector current combinations of different quarks. 
The generator $\lambda^0$ gives the  flavor singlet combination, $\lambda^3$ gives the flavor isovector (triplet) combination,  $\lambda^8$ gives the flavor hypercharge axial-vector (octet) combination  and $\lambda^{15}$ gives the flavor charmed combination. The respective axial-vector currents are expressed as 
\bea
A^{\mu,0}&=& \overline{u} (x)\gamma^\mu \gamma_5  u(x)+\overline{d} (x)\gamma^\mu \gamma_5  d(x)+\overline{s} (x)\gamma^\mu \gamma_5  s(x)+\overline{c} (x)\gamma^\mu \gamma_5  c(x)\,, \nonumber \\
A^{\mu,3}&=& \overline{u} (x)\gamma^\mu \gamma_5  u(x)-\overline{d} (x)\gamma^\mu \gamma_5  d(x)\,, \nonumber \\
A^{\mu,8}&=& \overline{u} (x)\gamma^\mu \gamma_5  u(x)+\overline{d} (x)\gamma^\mu \gamma_5  d(x)-2\overline{s} (x)\gamma^\mu \gamma_5  s(x)\,, \nonumber \\
A^{\mu,15}&=& \overline{u} (x)\gamma^\mu \gamma_5  u(x)+\overline{d} (x)\gamma^\mu \gamma_5  d(x)+\overline{s} (x)\gamma^\mu \gamma_5  s(x)-3\overline{c} (x)\gamma^\mu \gamma_5  c(x)\,. 
\eea
The axial-vector charges for all the members of the spin $\frac{1}{2}^+$ and spin $\frac{3}{2}^+$ baryon multiplet can now be calculated at zero momentum transfer. The different combinations can be represented in terms of the explicit spin polarizations of each baryon. We have
\bea
g^0_{B} &=&  \Delta u_{B}+\Delta d_{B}+\Delta s_{B} +\Delta c_{B}\,, \nonumber \\
g^3_{B} &=&  \Delta u_{ B}-\Delta d_{ B} \,, \nonumber \\
g^8_{B} &=&  \Delta u_{ B}+\Delta d_{B}-2\Delta s_{ B} \,, \nonumber \\
g^{15}_{B} &=&  \Delta u_{ B}+\Delta d_{B}+\Delta s_{ B}-3\Delta c_{ B} \,. \label{ggg}
\eea

\section{SU(4) chiral constituent quark model ($\chi$CQM)}
\label{cccm} In the $\chi$CQM, the basic process is the
internal emission of a Goldstone Boson (GB) by a constituent quark in the light and charmed baryons. This GB which consists of a 15-plet and a singlet further splits into a $q \bar q$ pair as \be q_{\pm}
\rightarrow {\rm GB}^{0}+q'_{\mp} \rightarrow (q \bar
q^{'})+q'_{\mp}\,.  \ee
This $q \bar q^{'} +q^{'}$ constitutes the
``quark sea'' \cite{cheng,hd,song,johan,hdcharm} and the effective  Lagrangian for the {\it
	SU}(4) $\chi$CQM describing interaction between
quarks and 16 GBs can be
expressed as \be {{\cal L}} = g_{15}{\bar \psi}\left( \Phi' \right)
{\psi} \,. \label{basic4} \ee Here $g_{15}$ is the coupling
constant,  $\psi$ is the quark field  \bea \psi = \left( \ba{c} u \\d
\\s\\c \ea \right)\,,\eea 
and $\Phi'$is the GB field expressed as 
	\bea  \left( \ba{ccccc} \frac{\pi^0}{\sqrt 2}
	+\beta\frac{\eta}{\sqrt 6}+\zeta\frac{\eta^{'}}{4\sqrt
		3}-\gamma\frac{\eta_c}{4} & \pi{^+} & \alpha K{^+} & \gamma
	\bar{D}^0\\ \pi^- & -\frac{\pi^0}{\sqrt 2} +\beta
	\frac{\eta}{\sqrt 6} +\zeta\frac{\eta^{'}}{4\sqrt 3}
	-\gamma\frac{\eta_c}{4}& \alpha K^0 & \gamma D^-\\ \alpha K^- &
	\alpha \bar{K}^0 &- \beta \frac{2\eta}{\sqrt 6} +
	\zeta\frac{\eta^{'}}{4\sqrt 3}- \gamma\frac{\eta_c}{4} & \gamma
	D^-_s\\ \gamma D^0  &\gamma D{^+}& \gamma D^+_s&-
	\zeta\frac{3\eta^{'}}{4\sqrt 3}+ \gamma\frac{3\eta_c}{4} \ea
	\right)\,. \nonumber\eea 
A parameter $a(=|g_{15}|^2)$ denoting the transition probability of chiral
fluctuation of the splitting  $u(d) \rightarrow d(u)+ \pi^{+(-)}$ can be introduced. 
The probabilities of transitions of $u(d) \rightarrow s +
K^{-(o)}$, $u(d,s) \rightarrow u(d,s) + \eta$, $u(d,s) \rightarrow
u(d,s) + \eta^{'}$ and  $u(d) \rightarrow c +\bar{D}^0(D^-)$ are then denoted by $a \alpha^2$, $a \beta^2$, $a \zeta^2$ and $a \gamma^2$ respectively.  The masses
of GBs are considered to be non-degenerate $(M_{\eta_{c}}> M_{\eta^{'}}>
M_{K,\eta}>M_{\pi})$
and symmetry breaking is introduced by
considering $M_c>M_s>M_{u,d}$.

\section{Spin polarizations for spin$-{\frac{1}{2}}^+$ and spin$-{\frac{3}{2}}^+$ charmed baryon multiplets} \label{}

The spin structure of the spin $\frac{1}{2}^+$ and spin $\frac{3}{2}^+$ charmed baryon multiplets are defined as
\cite{cheng,hd,johan} \bea \widehat B^{\frac{1}{2}^+} &\equiv& \langle B^{\frac{1}{2}^+}(p', J_z')|{\cal N}|B^{\frac{1}{2}^+}(p, J_z)
\rangle\,,  \nonumber \\ 
\widehat B^{\frac{3}{2}^+} &\equiv& \langle B^{\frac{3}{2}^+}(p', J_z')|{\cal N}|B^{\frac{3}{2}^+}(p, J_z)
\rangle\,, \label{bnb} \eea where $|B^{\frac{1}{2}^+}\rangle$ and $|B^{\frac{3}{2}^+}\rangle$ are the  wave
functions for the spin $\frac{1}{2}^+$ and spin $\frac{3}{2}^+$ baryon multiplets respectively. The number operator ${\cal N}$ defined in terms of the number of $q_{\pm}$ quarks ($n_{q_{\pm}}$) can be defined as  \be
{\cal N}=n_{u_{+}}u_{+}+ n_{u_{-}}u_{-} + n_{d_{+}}d_{+} +
n_{d_{-}}d_{-} + n_{s_{+}}s_{+} + n_{s_{-}}s_{-} + n_{c_{+}}c_{+}
+ n_{c_{-}}c_{-}\,, \label{number} \ee 
and can be used to calculate the spin polarizations ($\Delta q= q_{+}- q_{-}$) for a given baryon. 
For each constituent quark the substitution  made is \be q_{\pm}\rightarrow \sum P_{[q, ~ GB]} q_{\pm}+ |\psi(q_{\pm})|^2\,,
\label{qpq} \ee
where $\sum P_{[q, ~ GB]}$ is the probability of emission of
GBs from a $q$ quark and $|\psi(q_{\pm})|^2$ is the transition probability of
fluctuation of every constituent $q_{\pm}$ quark  \cite{hdcharm}.
We have \bea P_{[u, ~ GB]} &=&
a\left( \frac{3}{2} +\alpha^2 + \frac{\beta^2}{6} +
\frac{\zeta^2}{48} + \gamma^2 \right) \,, \nonumber
\\ P_{[d, ~ GB]}  &=& a\left( \frac{3}{2} +\alpha^2
+ \frac{\beta^2}{6} + \frac{\zeta^2}{48} + \gamma^2 \right) \,,\nonumber \\
P_{[s, ~GB]}  &=& a\left( 2 \alpha^2 + \frac{2\beta^2}{3} +
\frac{\zeta^2}{48} + \gamma^2 \right)\,,  \nonumber\\
P_{[c, ~GB]} &=& a \left( \frac{3\zeta^2}{16} +\frac{57
	\gamma^2}{16} \right) \,, \label{putc} \eea and 
\bea |\psi(u_{\pm})|^2  &=& a\left(\frac{1}{2} +
\frac{\beta^2}{6} + \frac{\zeta^2}{48} + \frac{\gamma^2}{16} \right)
u_{\mp} + a d_{\mp} + a \alpha^2 s_{\mp}+ a \gamma^2 c_{\mp}\,,
\label{psiupc}  \nonumber \\
|\psi(d_{\pm})|^2 
&=& a u_{\mp}+ a \left(\frac{1}{2} + \frac{\beta^2}{6} +
\frac{\zeta^2}{48} + \frac{\gamma^2}{16} \right)d_{\mp} + a \alpha^2
s_{\mp}+ a \gamma^2 c_{\mp}\,, \label{psidpc}  \nonumber \\
|\psi(s_{\pm})|^2 
&=& a \alpha^2 u_{\mp}+ a \alpha^2 d_{\mp} + a \left(\frac{2}{3}
\beta^2 + \frac{\zeta^2}{48} + \frac{\gamma^2}{16} \right)s_{\mp} +
+ a \gamma^2 c_{\mp} \,,\label{psispc}  \nonumber \\
|\psi(c_{\pm})|^2 
&=& a \gamma^2 u_{\mp} + a \gamma^2 d_{\mp} + a \gamma^2 s_{\mp} + a
\left( \frac{3 \zeta^2}{16} + \frac{9 \gamma^2}{16} \right) c_{\mp}
\,. \label{psicpc} \eea 

The spin polarizations for each baryon can be computed using Eqs. (\ref{bnb}) and (\ref{number}). In general, the wave function for the three-quark system made from any of
the $u$, $d$, $s$, or $c$ quarks is given as a product of $\phi \chi \psi$, with $\phi$ being the
flavor part, $\chi$ the spin part and $\psi$  the
spatial part of the wave function. The  wave
functions for the spin $\frac{1}{2}^+$ baryon multiplet is expressed as 

\bea |B^{\frac{1}{2}^+}\rangle \simeq |120,
{^2}20_M\rangle_{N=0}&=& \frac{1}{\sqrt 2}(\chi^{'}
\phi^{'} + \chi^{''} \phi^{''}) \psi^{s}(0^+) \,.
\eea
The effect of configuration mixing generated by the
spin-spin interactions \cite{dgg,isgur,yaouanc,mmgupta}  can be
incorporated in the complete wave function  for the case of spin $\frac{1}{2}^+$ baryon multiplet and we have \bea |B^{\frac{1}{2}^+}\rangle &=& {\cos} \theta |120,
{^2}20_M\rangle_{N=0} + {\sin} \theta|168,{^2}20_M\rangle_{N=2}\,,
\label{config} \eea
 where \be |168, {^2}20_M \rangle_{N=2} =\frac{1}{2}(( \phi^{'}\chi^{''} + \phi^{''} \chi^{'})\psi^{'}(0^+) +(\phi^{'} \chi^{'} - \phi^{''} \psi^{''}) \psi^{''}(0^+)).\ee
The spin $\frac{3}{2}^+$ baryon multiplet can be expressed as 
\bea
|B^{\frac{3}{2}^+}\rangle\simeq |120, {^4}20_S{\rangle_{N=0}}&=& \chi^{s}
\phi^s \psi^s(0^+) \,. \label{3/2}\eea 
The content of the ${\it SU}(4) \otimes {\it SU}(2)=SU(8)$ flavor and spin multiplets is given as \bea 120
&\supset&{^4}20_S+{^2}20_M \,,\nonumber\\ 168
&\supset&{^2}20_S+{^4}20_M+{^2}20_M+{^2}\bar 4 \,,\nonumber\\ 56
&\supset& {^2}\bar 4+{^2}20_M \,. \eea 
 For the details of the
definition of flavor, spin and spatial part of the wave function we refer the
reader to Ref. \cite{nscharm,yaoubook}.

The spin part of wavefunction $\chi$ is expressed as
\bea \chi^{s} &=&  \uparrow \uparrow \uparrow\,, \nonumber \\
\chi^{'} &= & \frac{1}{\sqrt 2}(\uparrow \downarrow \uparrow
-\downarrow \uparrow \uparrow)\,, \nonumber \\ \chi^{''} &=&
\frac{1}{\sqrt 6} (2\uparrow \uparrow \downarrow -\uparrow
\downarrow \uparrow -\downarrow \uparrow \uparrow)\,. \eea

The explicit forms of the flavor wavefunctions
$\phi^{'}$ and $\phi^{''}$ for the spin
${\frac{1}{2}}^+$  baryons of the type
$B(xxy)$ are \bea \phi^{'}_B &=& \frac{1}{\sqrt 2}(xyx-yxx)\,,
\nonumber \\ \phi^{''}_B &=& \frac{1}{\sqrt 6}(2xxy-xyx-yxx)\,, \eea
where $x$, $y$, and $z$ correspond to any of the $u$, $d$, $s$ and $c$
quarks. For the case of the baryons of the type $B(xyz)$, the
wavefunctions are given as \bea \phi^{'} &=&
\frac{1}{2 \sqrt 3}(xzy+zyx-zxy-yzx-2xyz-2yxz)\,, \nonumber \\
&&{\rm or} \nonumber \\
\phi^{'}&=& \frac{1}{2}(zxy+zyx-xzy-yzx)\,, 
\eea
\bea
\phi^{''} &=& \frac{1}{2}(zxy+xzy-zyx-yzx)\,, \nonumber \\
&&{\rm or} \nonumber \\
\phi^{''} &=&\frac{1}{2 \sqrt
	3}(zyx+zxy+xzy+yzx-2xyz-2yxz)\,. \eea The flavor wavefunctions
$\phi^{s}$ for the spin
${\frac{3}{2}}^+$  baryons of the types $B^*(xxx)$,
$B^*(xxy)$, and $B^*(xyz)$, respectively, are expressed as \bea
\phi^{s}_{B^*} &=& xxx\,, \nonumber \\
\phi^{s}_{B^*} &=& \frac{1}{\sqrt 3}(xxy + xyx + yxx)\,, \nonumber \\
\phi^{s}_{B^*} &=& \frac{1}{\sqrt 6}(xyz + xzy + yxz + yzx + zxy +
zyx)\,. \eea 
The explicit flavor wave functions for the spin ${\frac{3}{2}}^+$
and spin ${\frac{1}{2}}^+$ baryons  are given in Tables \ref{wf3-2} and \ref{wf1-2}. Further details and definitions of the spatial wave
functions ($\psi^{s}, \psi^{'}, \psi^{''})$ as well as the
definitions of the overlap integrals are discussed in Ref. \cite{yaoubook}.

\begin{sidewaystable}
	\tabcolsep 1.0mm {\renewcommand{\arraystretch}{1.0}
		\begin{center}
			\begin{tabular}{|cccc|}
				\hline \hline \multicolumn{2}{|c}{{ Spin~ ${\frac{1}{2}}^+$~ Baryons}} &$ \phi{'}$ &$ \phi{''}$\\
				\hline \hline (8,0)& $p$ & $\frac{1}{\sqrt 2}(udu-duu)$&$ \frac{1}{\sqrt
					6}(2uud-udu-duu) $ \\ &$ n$ &$\frac{1}{\sqrt
					2}(udd-dud)$&$\frac{1}{\sqrt 6}(dud+udd-2ddu)$\\
				&$\Sigma^{+}$ &$\frac{1}{\sqrt 2}(usu-suu)$ &$ \frac{1}{\sqrt
					6}(2uus-suu-usu)$\\&$\Sigma^{0}$ &$
				\frac{1}{2}(sud+sdu-usd-dsu)$ &$ \frac{1}{2\sqrt
					3}(sdu+dsu+sud+usd-2uds-2dus)$\\ &$\Sigma^-$ &$\frac{1}{\sqrt
					2}(sdd-dsd)$ &$ \frac{1}{\sqrt 6}(2dds-sdd-dsd)$ \\ &$ \Lambda
				$ &$\frac{1}{2\sqrt 3}(2uds-2dus+sdu-dsu+usd-sud)$ & $
				\frac{1}{2}(sud+usd-sdu-dsu)$ \\ &$\Xi^0$ &$\frac{1}{\sqrt
					2}(sus-uss)$ & $\frac{1}{\sqrt 6}(sus+uss-2ssu)$ \\
				&$\Xi^-$ & $\frac{1}{\sqrt 2}(sds-dss)$ & $\frac{1}{\sqrt
					6}(sds+dss-2ssd)$ \\ \hline (6,1) & $\Sigma_{c}^{++}$&$
				\frac{1}{\sqrt 2}(cuu-ucu)$ &$\frac{1}{\sqrt
					6}(cuu+ucu-2uuc)$\\&$\Sigma_{c}^{+}$ &$\frac{1}{2}(cud+cdu-ucd-dcu)$
				&$\frac{1}{2\sqrt 3}(dcu+cdu+ucd+cud-2udc-2duc)$ \\ &$\Sigma_{c}^{0}$&$\frac{1}{\sqrt 2}(cdd-dcd)$ &
				$\frac{1}{\sqrt 6}(cdd+dcd-2ddc)$ \\ &$\Xi_{c}^{'+}$ &$\frac{1}{2}(cus+csu-ucs-scu)$ &$\frac{1}{
					2\sqrt 3}(ucs+cus+scu+csu-2usc-2suc)$ \\ &$\Xi_{c}^{'0}
				$ &$\frac{1}{2}(cds+csd-dcs-scd)$ &$\frac{1}{2\sqrt
					3}(dcs+cds+scd+csd-2dsc-2sdc)$
				\\&$\Omega_{c}^{0}$ &$\frac{1}{\sqrt 2}(css-scs)$ & $\frac{1}{\sqrt
					6}(scs+css-2ssc)$ \\ \hline $({\bar 3},
				1)$ &$\Lambda_{c}^{+}$ &$\frac{1}{2\sqrt 3}(2udc-2duc+cdu-dcu+ ucd-cud)$
				&$\frac{1}{2}(ucd+cud-dcu-cdu)$\\
				&$\Xi_{c}^{+}$ &$\frac{1}{2\sqrt 3}(2usc-2suc+csu-scu+
				ucs-cus)$ &$\frac{1}{2}(ucs+cus-scu-csu)$\\
				&$\Xi_{c}^{0}$ & $\frac{1}{2\sqrt
					3}(2dsc-2sdc+csd-scd+dcs-cds)$ &$\frac{1}{2}(dcs+cds-scd-csd)$
				\\ \hline(3,2)&$\Xi_{cc}^{++}$ &$\frac{1}{\sqrt 2}(ucc-cuc)$
				&$\frac{1}{\sqrt 6}(ucc+cuc-2ccu)$
				\\&$\Xi_{cc}^{+}$ &$\frac{1}{\sqrt 2}(dcc-cdc) $ & $\frac{1}{\sqrt
					6}(dcc+cdc-2ccd)$\\& $\Omega_{cc}^{+}$ &$\frac{1}{\sqrt
					2}(scc-csc)$ &$\frac{1}{\sqrt6}(scc+csc-2ccs)$ \\ \hline
				\hline
			\end{tabular}
	\caption{The explicit flavor wave functions for the spin ${\frac{3}{2}}^+$
	 baryons} \label{wf3-2}
		\end{center}
	}
\end{sidewaystable}

\begin{table}
	\tabcolsep 2.5mm {\renewcommand{\arraystretch}{1.3}
		
		\begin{tabular}{|ccc|} \hline \hline
			\multicolumn{2}{|c}{{ Spin~ ${\frac{3}{2}}^+$~ Baryons}}&$ \phi^s $\\
			\hline \hline (10,0) &$ \Delta^{++} $&$uuu$\\
			&$\Delta^{+}$&$\frac{1}{\sqrt 3}(uud+udu+duu)$\\
			&$\Delta^{0}$&$\frac{1}{\sqrt 3}(udd+ddu+dud)$\\
			&$\Delta^{-}$&$ddd $\\ &$\Sigma^{*+}$&$\frac{1}{\sqrt
				3}(uus+suu+usu)$\\ &$\Sigma^{*-}$&$\frac{1}{\sqrt
				3}(dds+dsd+sdd)$\\ &$\Sigma^{*0}$&$\frac{1}{\sqrt
				6}(sdu+sud+usd+dsu+dus+uds)$\\ &$\Xi^{*0} $&$\frac{1}{\sqrt
				3}(ssu+sus+uss)$\\ &$\Xi^{*-}$&$\frac{1}{\sqrt
				3}(ssd+sds+dss)$\\ &$\Omega^{-}$&$ sss  $\\\hline
			(6,1) &$\Sigma_{c}^{*++}$&$ \frac{1}{\sqrt 3}(uuc+ucu+cuu)$\\
			&$\Sigma_{c}^{*+}$&$ \frac{1}{\sqrt
				6}(udc+dcu+cud+cdu+duc+ucd)$\\ &$\Sigma_{c}^{*0}$&$
			\frac{1}{\sqrt 3}(ddc+dcd+cdd)$\\ &$\Xi_{c}^{*+}$&$
			\frac{1}{\sqrt 6}(usc+scu+cus+csu+suc+ucs)$\\
			&$\Xi_{c}^{*0}$&$ \frac{1}{\sqrt
				6}(dsc+scd+cds+csd+dsc+scd)$\\
			&$\Omega_{c}^{*0}$&$
			\frac{1}{\sqrt 3}(ssc+scs+css)$\\\hline
			(3,2) &$\Xi_{cc}^{*++}$&$\frac{1}{\sqrt 3}(ucc+cuc+ccu)$\\
			&$\Xi_{cc}^{*+}$&$\frac{1}{\sqrt 3}(dcc+cdc+ccd)$\\
			&$\Omega_{cc}^{*+}$&$ \frac{1}{\sqrt 3}(scc+csc+ccs)$\\\hline
			(1,3)&$\Omega_{ccc}^{*++}$&$ccc$ \\ \hline \hline
		\end{tabular} \caption{The explicit flavor wave functions for the spin ${\frac{1}{2}}^+$
		baryons} \label{wf1-2}
		
	}
\end{table}

\subsection{Spin polarizations for spin
	${\frac{1}{2}}^+$ charmed baryon multiplets}

 Using Eqs. (\ref{bnb}), (\ref{number}) and (\ref{qpq}) we can compute the spin polarizations for the spin $\frac{1}{2}^+$ baryon multiplet employing the corresponding wave function from Eq. (\ref{config}). In Tables \ref{table_baryon_octet_spin_half}-\ref{table_double_charmed_spin_half} we have presented the explicit expressions for the spin $\frac{1}{2}^+$ mixed symmetry 20-plet with the light baryon octet (8,0), singly charmed sextet (6,1) as well as  anti-triplet ($\bar{3}$,2) and triplet (3,2) with doubly charmed baryons.  In Table \ref{table_baryon_octet_spin_half} we have presented the explicit expressions for the octet of baryons containing only $u$, $d$ and $s$ quarks.  In Table \ref{table_single_charmed_spin_half} the explicit expressions for the singly charmed sextet baryons have been presented and in Table \ref{table_double_charmed_spin_half} we have given the doubly charmed anti-triplet and triplet baryons.

\subsection{Spin polarizations for spin
	${\frac{3}{2}}^+$ charmed baryon multiplets}

In Tables \ref{table_decuplet_3_2}-\ref{table_double_charm_3_2} we have presented the explicit expressions for the spin $\frac{3}{2}^+$  symmetric 20-plet with the light baryon decuplet (10,0), singly charmed sextet (6,1), doubly charmed triplet (3,2) and triply charmed singlet (1,3).  In Table \ref{table_decuplet_3_2} we have presented the explicit expressions for the decuplet of baryons containing only $u$, $d$ and $s$ quarks.  In Table \ref{table_single_charm_3_2} the explicit expressions for the singly charmed sextet baryons have been presented and in Table \ref{table_double_charm_3_2} we have given the doubly charmed triplet and the triply charmed singlet baryons.


\section{Input parameters} \label{input4}
In order to calculate the numerical values of the axial-vector charges, we now discuss the various input parameters used in the present calculations. We have to first fix the symmetry breaking parameters of the SU(4) $\chi$CQM. The  transition probabilities give the extent of contribution from the non-constituent quarks to the structure of the baryon. We have the  probabilities of fluctuations of a constituent
quark into $\pi$, $K$, $\eta$, $\eta^{'}$, $\eta_c$ represented respectively by $a$, $a\alpha^2$, $a \beta^2$, $a \zeta^2$, $a \gamma^2$. Since these probabilities depend on the mass difference, the probability of emitting a heavier meson such as $D-$meson should be less as compared to the probability of emitting a lighter meson such as $K$, $\eta$, $\eta{'}$ and  $\pi$.
These parameters have already been fixed and applied successfully in the context of the calculations of  spin and flavor distribution
 functions as well as magnetic moments of spin$-{\frac{1}{2}}^+$ and spin$-{\frac{3}{2}}^+$ charmed baryons including
 their radiative decays \cite{hdcharm,nscharm}. 
 Another parameter used here is the mixing angle $\phi$ for the case of spin$-{\frac{1}{2}}^+$ charmed baryon multiplets. It is fixed by fitting neutron
 charge radius \cite{yaouanc}.   In the present work, we will use the set of parameters obtained by a fine grained analysis  and are listed as
 \be \phi =20^o\,,~~a=0.12\,,~~ a \alpha^2 \simeq a \beta^2=
 0.0243\,,~~ a \zeta^2 = 0.0053 \,, ~~ {\rm and} ~~
 a \gamma^2=0.0014\,.\ee
 
\section{Spin polarizations and axial-vector charges for spin$-{\frac{1}{2}}^+$ and spin$-{\frac{3}{2}}^+$ charmed baryon multiplets}

Using the above set of input parameters we have presented in Tables \ref{QuarkPolar_Spin_1by2_Octet}-\ref{QuarkPolar_Spin_3by2_triplet_and_singlet} the results of spin polarizations and axial-vector charges for charmed baryons in the spin$-{\frac{1}{2}}^+$ and spin$-{\frac{3}{2}}^+$ multiplets to understand their quantitative contributions. 
In Table \ref{QuarkPolar_Spin_1by2_Octet} we have presented the results for the spin $\frac{1}{2}^+$ mixed symmetry 20-plet with the light baryon octet (8,0), in Table \ref{QuarkPolar_Spin_1by2_Sextet} the singly charmed sextet (6,1) and in Table \ref{QuarkPolar_Spin_1by2_triplet} the anti-triplet ($\bar{3}$,2) and triplet (3,2) with doubly charmed baryons. Further, in Table \ref{QuarkPolar_Spin_3by2_decuplet} we have presented the results for the spin $\frac{3}{2}^+$  symmetric 20-plet with the light baryon decuplet (10,0), in Table \ref{QuarkPolar_Spin_3by2_Sextet} the singly charmed sextet (6,1) and in Table \ref{QuarkPolar_Spin_3by2_triplet_and_singlet} the doubly charmed triplet (3,2) and triply charmed singlet (1,3).

For the case of spin$-{\frac{1}{2}}^+$ charmed baryons, we observe the following relations between the isospin partners with the interchange of $u \longleftrightarrow d$
\bea
\ba{lllll}
\Delta u_{p}=\Delta d_{n}, & g^0_{p}=g^0_{n}, & g^3_{p}=-g^3_{n}, &  g^8_{p}=g^8_{n}, & g^{15}_{p}=g^{15}_{n}, \\
\Delta u_{\Sigma^+}=\Delta d_{\Sigma^-}, & g^0_{\Sigma^+}=g^0_{\Sigma^-}, & g^3_{\Sigma^+}=-g^3_{\Sigma^-}, &  g^8_{\Sigma^+}=g^8_{\Sigma^-}, & g^{15}_{\Sigma^+}=g^{15}_{\Sigma^-},\\
\Delta u_{\Xi^0}=\Delta d_{\Xi^-}, & g^0_{\Xi^0}=g^0_{\Xi^-}, & g^3_{\Xi^0}=-g^3_{\Xi^-}, &  g^8_{\Xi^0}=g^8_{\Xi^-}, & g^{15}_{\Xi^0}=g^{15}_{\Xi^-},\\
\Delta u_{\Sigma_c^{++}}=\Delta d_{\Sigma_c^0}, & g^0_{\Sigma_c^{++}}=g^0_{\Sigma_c^0}, & g^3_{\Sigma_c^{++}}=-g^3_{\Sigma_c^0}, &  g^8_{\Sigma_c^{++}}=g^8_{\Sigma_c^0}, & g^{15}_{\Sigma_c^{++}}=g^{15}_{\Sigma_c^0},\\
\Delta u_{\Xi_c^{\prime +}}=\Delta d_{\Xi_c^{\prime 0}}, & g^0_{\Xi_c^{\prime +}}=g^0_{\Xi_c^{\prime 0}}, & g^3_{\Xi_c^{\prime +}}=-g^3_{\Xi_c^{\prime 0}}, &  g^8_{\Xi_c^{\prime +}}=g^8_{\Xi_c^{\prime 0}}, & g^{15}_{\Xi_c^{\prime +}}=g^{15}_{\Xi_c^{\prime 0}},\\
\Delta u_{\Xi_c^{+}}=\Delta d_{\Xi_c^{0}}, & g^0_{\Xi_c^{+}}=g^0_{\Xi_c^{0}}, & g^3_{\Xi_c^{+}}=-g^3_{\Xi_c^{ 0}}, &  g^8_{\Xi_c^{+}}=g^8_{\Xi_c^{0}}, & g^{15}_{\Xi_c^{+}}=g^{15}_{\Xi_c^{0}},\\
\Delta u_{\Xi_{cc}^{++}}=\Delta d_{\Xi_{cc}^{+}}, & g^0_{\Xi_{cc}^{++}}=g^0_{\Xi_{cc}^{+}}, & g^3_{\Xi_{cc}^{++}}=-g^3_{\Xi_{cc}^{+}}, &  g^8_{\Xi_{cc}^{++}}=g^8_{\Xi_{cc}^{+}}, & g^{15}_{\Xi_{cc}^{++}}=g^{15}_{\Xi_{cc}^{+}}.
\ea
\eea
We also have the same contribution of the strange quark polarizations in the light and charmed baryons lying in the same multiplets between each of the octet, sextet, anti-triplet and triplet. We have
\bea
\ba{l}
\Delta s_{p}=\Delta s_{n}\,, \nonumber \\
\Delta s_{\Sigma^+}= \Delta s_{\Sigma^0}=\Delta s_{\Sigma^-}\,, \nonumber \\
\Delta s_{\Xi^0}=\Delta s_{\Xi^-}\,, \nonumber \\
\Delta s_{\Sigma_c^{++}}=\Delta s_{\Sigma_c^+}=\Delta s_{\Sigma_c^0}\,, \nonumber \\
\Delta s_{\Xi_c^{\prime +}}=\Delta s_{\Xi_c^{\prime 0}}\,, \nonumber \\
\Delta s_{\Xi_c^{+}}=\Delta s_{\Xi_c^{0}}\,, \nonumber \\
\Delta s_{\Xi_{cc}^{++}}=\Delta s_{\Xi_{cc}^{+}}\,.
\ea
\eea
Further, the charm quark polarization contributes equally for the entire octet, sextet, anti-triplet and triplet baryons since the number of charmed quarks are same in the constituent structure. We have
\bea
\ba{l}
\Delta c_{p}=\Delta c_{n}=\Delta c_{\Sigma^+}= \Delta c_{\Sigma^0}=\Delta c_{\Sigma^-}=\Delta c_{\Lambda}=\Delta c_{\Xi^0}=\Delta c_{\Xi^-}\,, \nonumber \\
\Delta c_{\Sigma_c^{++}}=\Delta c_{\Sigma_c^+}=\Delta c_{\Sigma_c^0}=\Delta c_{\Xi_c^{\prime +}}=\Delta c_{\Xi_c^{\prime 0}}=\Delta c_{\Omega_c^{0}}\,, \nonumber \\
\Delta c_{\Lambda_c^{+}}=\Delta c_{\Xi_c^{+}}=\Delta c_{\Xi_c^{0}}\,, \nonumber \\
\Delta c_{\Xi_{cc}^{++}}=\Delta c_{\Xi_{cc}^{+}}=\Delta c_{\Omega_{cc}^{+}}\,.
\ea
\eea

Similarly, for the case of spin$-{\frac{3}{2}}^+$ light and charmed baryons we observe the following relations between the isospin partners with the interchange of $u \longleftrightarrow d$
\bea
\ba{lcccc}
\Delta u_{\Delta^{++}}=\Delta d_{\Delta^-}, & g^0_{\Delta^{++}}=g^0_{\Delta^-}, & g^3_{\Delta^{++}}=-g^3_{\Delta^-}, &  g^8_{\Delta^{++}}=g^8_{\Delta^-}, & g^{15}_{\Delta^{++}}=g^{15}_{\Delta^-}, \\
\Delta u_{\Delta^{+}}=\Delta d_{\Delta^{0}}, & g^0_{\Delta^{+}}=g^0_{\Delta^{0}}, & g^3_{\Delta^{+}}=-g^3_{\Delta^{0}}, &  g^8_{\Delta^{+}}=g^8_{\Delta^{0}}, & g^{15}_{\Delta^{+}}=g^{15}_{\Delta^{0}},\\
\Delta u_{\Sigma^{*+}}=\Delta d_{\Sigma^{*-}}, & g^0_{\Sigma^{*+}}=g^0_{\Sigma^{*0}}=g^0_{\Sigma^{*-}}, & g^3_{\Sigma^{*+}}=-g^3_{\Sigma^{*-}}, &  g^8_{\Sigma^{*+}}=g^8_{\Sigma^{*0}}=g^8_{\Sigma^{*-}}, & g^{15}_{\Sigma^{*+}}=g^{15}_{\Sigma^{*0}}=g^{15}_{\Sigma^{*-}},\\
\Delta u_{\Xi^{*0}}=\Delta d_{\Xi^{*-}}, & g^0_{\Xi^{*0}}=g^0_{\Xi^{*-}}, & g^3_{\Xi^{*0}}=-g^3_{\Xi^{*-}}, &  g^8_{\Xi^{*0}}=g^8_{\Xi^{*-}}, & g^{15}_{\Xi^{*0}}=g^{15}_{\Xi^{*-}},\\
\Delta u_{\Sigma_c^{*++}}=\Delta d_{\Sigma_c^{*0}}, & g^0_{\Sigma_c^{*++}}=g^0_{\Sigma_c^{*+}}=g^0_{\Sigma_c^{*0}}, & g^3_{\Sigma_c^{*++}}=-g^3_{\Sigma_c^{*0}}, &  g^8_{\Sigma_c^{*++}}=g^8_{\Sigma_c^{*+}}=g^8_{\Sigma_c^{*0}}, & g^{15}_{\Sigma_c^{*++}}=g^{15}_{\Sigma_c^{*+}}=g^{15}_{\Sigma_c^{*0}},\\
\Delta u_{\Xi_c^{*+}}=\Delta d_{\Xi_c^{*0}}, & g^0_{\Xi_c^{*+}}=g^0_{\Xi_c^{*0}}, & g^3_{\Xi_c^{*+}}=-g^3_{\Xi_c^{* 0}}, &  g^8_{\Xi_c^{*+}}=g^8_{\Xi_c^{*0}}, & g^{15}_{\Xi_c^{*+}}=g^{15}_{\Xi_c^{*0}},\\
 & g^0_{\Xi_{cc}^{*++}}=g^0_{\Xi_{cc}^{*+}}, &  &  g^8_{\Xi_{cc}^{*++}}=g^8_{\Xi_{cc}^{*+}}, & g^{15}_{\Xi_{cc}^{*++}}=g^{15}_{\Xi_{cc}^{*+}}.
\ea
\eea
The contribution of the strange quark polarizations is again same in the multiplets between each of the decuplet, sextet and triplet of the baryons. We have
\bea
\ba{l}
\Delta s_{\Delta^{++}}=\Delta s_{\Delta^+}=\Delta s_{\Delta^{0}}=\Delta s_{\Delta^-}\,, \nonumber \\
\Delta s_{\Sigma^{*+}}= \Delta s_{\Sigma^{*0}}=\Delta s_{\Sigma^{*-}}\,, \nonumber \\
\Delta s_{\Xi^{*0}}=\Delta s_{\Xi^{*-}}\,, \nonumber \\
\Delta s_{\Sigma_c^{*++}}=\Delta s_{\Sigma_c^{*+}}=\Delta s_{\Sigma_c^{*0}}\,, \nonumber \\
\Delta s_{\Xi_c^{*+}}=\Delta s_{\Xi_c^{*0}}\,, \nonumber \\
\Delta s_{\Xi_{cc}^{*++}}=\Delta s_{\Xi_{cc}^{*+}}\,.
\ea
\eea
The charm quark polarization gives the same contribution for the entire decuplet, sextet and triplet baryons and we have
\bea
\ba{l}
\Delta c_{\Delta^{++}}=\Delta c_{\Delta^+}=\Delta c_{\Delta^{0}}=\Delta c_{\Delta^-}=\Delta c_{\Sigma^{*+}}= \Delta c_{\Sigma^{*0}}=\Delta c_{\Sigma^{*-}}=\Delta c_{\Xi^{*0}}=\Delta c_{\Xi^{*-}}=\Delta c_{\Omega^{-}}\,, \nonumber \\
\Delta c_{\Sigma_c^{*++}}=\Delta c_{\Sigma_c^{*+}}=\Delta c_{\Sigma_c^{*0}}=\Delta c_{\Xi_c^{*+}}=\Delta c_{\Xi_c^{*0}}=\Delta c_{\Omega_c^{*0}}\,, \nonumber \\
\Delta c_{\Xi_{cc}^{*++}}=\Delta c_{\Xi_{cc}^{*+}}\,.
\ea
\eea

The magnitude of charm  quark polarization for different spin$-{\frac{1}{2}}^+$  and spin$-{\frac{3}{2}}^+$ light and charmed baryon multiplets is directly related to the number of charm quarks in the baryon and this observation is reflecting from the results as well. The value remains same within the baryons in the same multiplets but varies between the different multiplets having distinct values of hypercharge. We have
\bea
\ba{l}
\Delta c_{(8,0)}<\Delta c_{(6,1)}<\Delta c_{(\bar{3},2)}<\Delta c_{(3,2)}\,, \nonumber \\
\Delta c_{(10,0)}<\Delta c_{(6,1)}<\Delta c_{(3,2)}<\Delta c_{(1,3)}\,.
\ea
\eea

\section{$Q^2$ evolution and axial-vector charges for quarks}

The $Q^2$ evolution of the axial-vector charges of the light baryons has been experimentally investigated from the quasi elastic neutrino scattering \cite{antineutrino1,antineutrino2} and from the pion electroproduction \cite{pion-electro} in the past few years. These experiments have explored the $Q^2$ dependence of axial-vector charges but there is still more to understand in regard to the missing spin part of the nucleon. Some parametrization schemes have been adopted in literature for axial charges. For example, 
in  Ref. \cite{Bodek2008}, a parameterization scheme constrained by   quark–hadron duality is proposed for vector and axial-vector nucleon form factors.  In lattice QCD calculations $z-$ expansion parameterization is employed to study the $Q^2$ dependence of axial-form factors \cite{Jang2020}. The dipole for of parametrization conventionally used in the present work is most widely used parametrization  not only for axial form factors but also for the electric and magnetic form factors of baryons. The dipole form of parametrization used to analyze the axial-vector charges is given as 
\be
G^j_{B}(Q^2)=\frac{g^j_{B}(0)}{\left( 1+\frac{Q^2}{M_{A}^2}\right)^2}, \label{q2-dep}\ee
where $g_A^0(0)$, $g_A^3(0)$, $g_A^8(0)$ and $g_A^{15}(0)$ are the axial-vector coupling constants at zero momentum transfer. 
In Eq. (\ref{q2-dep}) $M_A$ is the axial mass. Neutrino-nuclei scattering experiments play a significant role in extracting the value of the axial mass parameter $M_A$. The value of $M_A$ extracted from measurement of
flux-averaged neutral-current elastic differential cross section at MiniBooNE experiment is $M_A = 1.39 \pm 0.11$ GeV at $Q^2 = 0$ \cite{miniboone}. From  MINOS \cite{adamson2015} and T2K \cite{abe2016} experiments, which involve the charged current quasi-elastic scattering from $^{56}{Fe}$ and $^{12}{C}$ respectively, the values of $M_A$ are found to be $1.23^{+0.13}_{-0.09}$ and $1.26^{+0.21}_{-0.18}$ GeV. In the present paper we use the value, $M_A = 1.10^{+0.13}_{-0.15}$ GeV, which has been extracted by employing
NuWro Monte Carlo generator method ($np$--$nh$ contribution) to the MiniBooNE experiment data \cite{golan2013}. In general, the  axial mass can be taken as free parameter and adjusted to experiment \cite{buchmann-axial-mass}. However, the experimental data being available only for the nucleon axial coupling constants, the same value of the axial mass for all the baryons has been used. The axial masses corresponding to the other light and charmed baryons are expected to be larger than that of the nucleon which will in turn lead to slightly larger values of the axial-vector charges in magnitude. The overall behavior of the form factors however will not be affected by this change.

In order to understand the internal structure of baryons and the dynamics of quarks inside them the explicit contributions of the constituent quarks in the baryons can also be studied. The constituent quarks can be viewed as spatially extended particles \cite{buchmann-axial-mass,q2range}, therefore the explicit quark axial-vector charges can be calculated. They can further be expressed in terms of the  singlet, triplet, octet and charmed combinations of the axial-vector charges as follows
\bea
G_{B}^u &=& \frac{1}{4} G_{B}^{0} + \frac{1}{2} G_{B}^{3} +\frac{1}{6} G_{B}^{8} + \frac{1}{12} G_{B}^{15}, \nonumber\\
G_{B}^d &=& \frac{1}{4} G_{B}^{0} - \frac{1}{2} G_{B}^{3} +\frac{1}{6} G_{B}^{8} + \frac{1}{12}Gg_{B}^{15}, \nonumber\\
G_{B}^s &=& \frac{1}{4} G_{B}^{0} - \frac{1}{3} G_{B}^{8} + \frac{1}{12} G_{B}^{15}, \nonumber\\
G_{B}^c &=& \frac{3}{4} G_{B}^{0} - \frac{1}{12} G_{B}^{15}. \nonumber\\
\eea

\begin{figure}
	\includegraphics {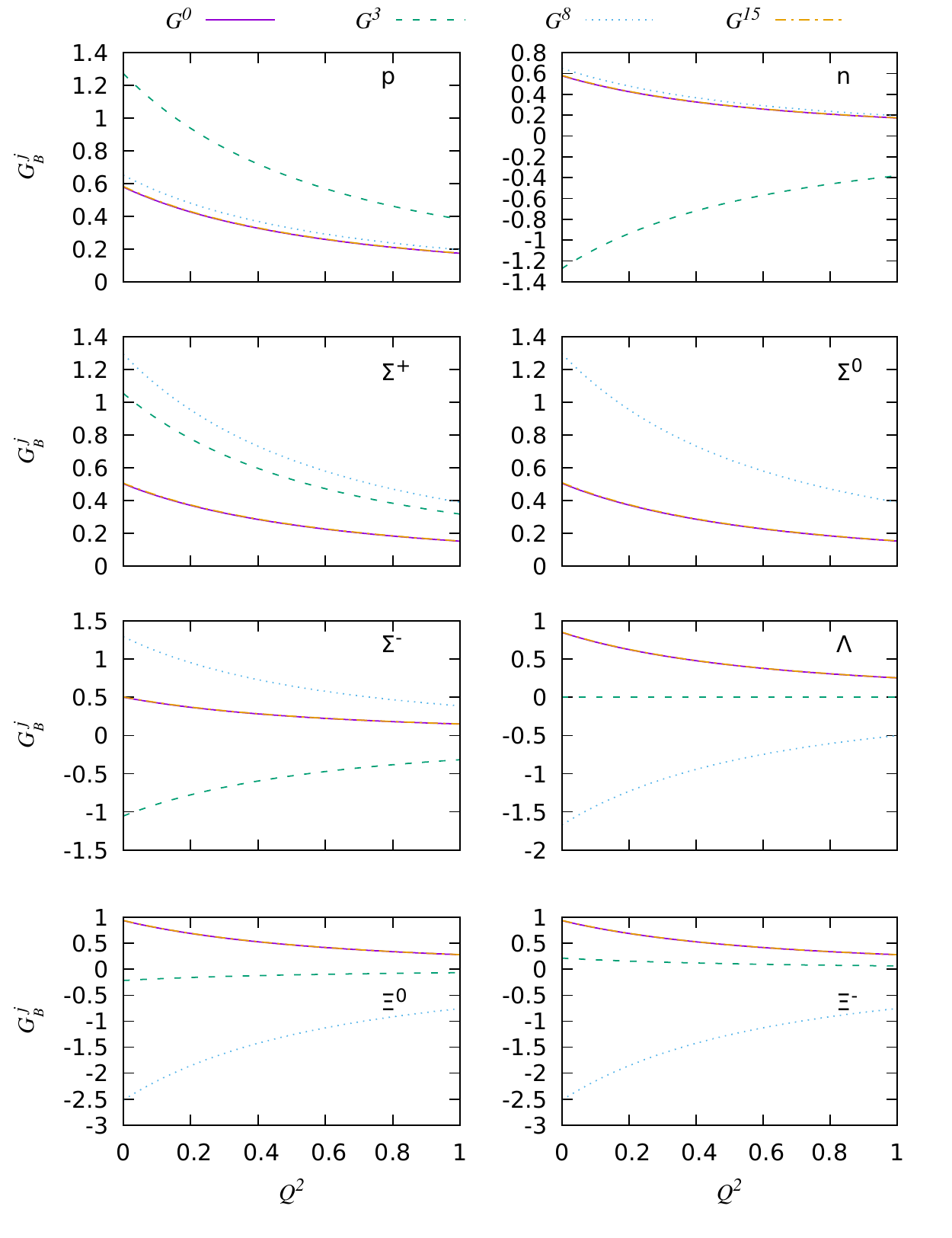}
	\caption{(color online). The axial-vector charges for  the spin $\frac{1}{2}^+$ mixed symmetry 20-plet with the light baryon octet (8,0) plotted as function of  $Q^2$ (in units of GeV$^2$).}
	\label{Spin_1by2_Octet}
\end{figure}

\begin{figure}
	\includegraphics {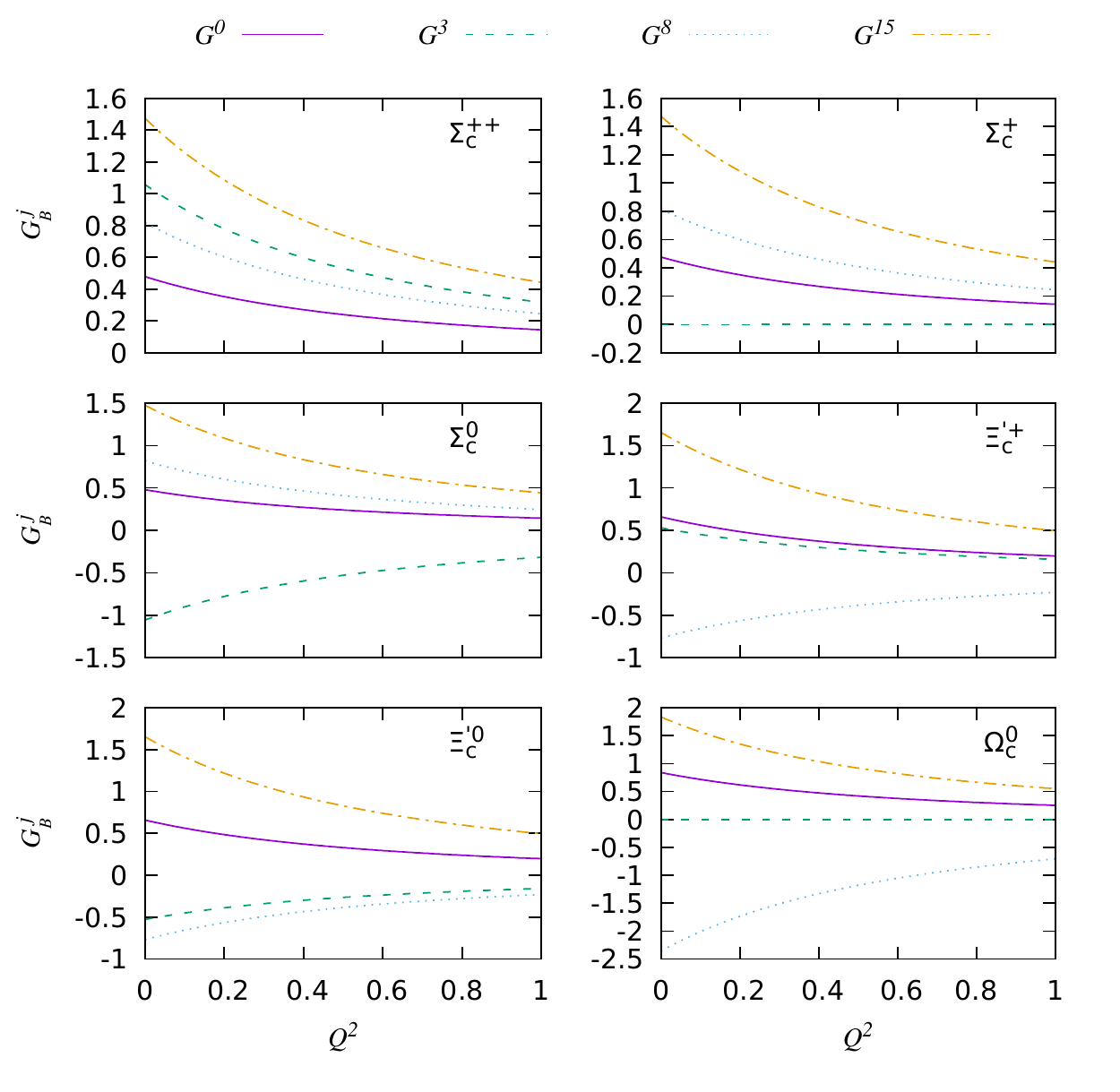}
	\caption{(color online). The axial-vector charges for  the spin $\frac{1}{2}^+$ mixed symmetry 20-plet with  the singly charmed sextet (6,1)  plotted as function of  $Q^2$ (in units of GeV$^2$).}
	\label{Spin_1by2_Sextet}
\end{figure}

\begin{figure}
	\includegraphics {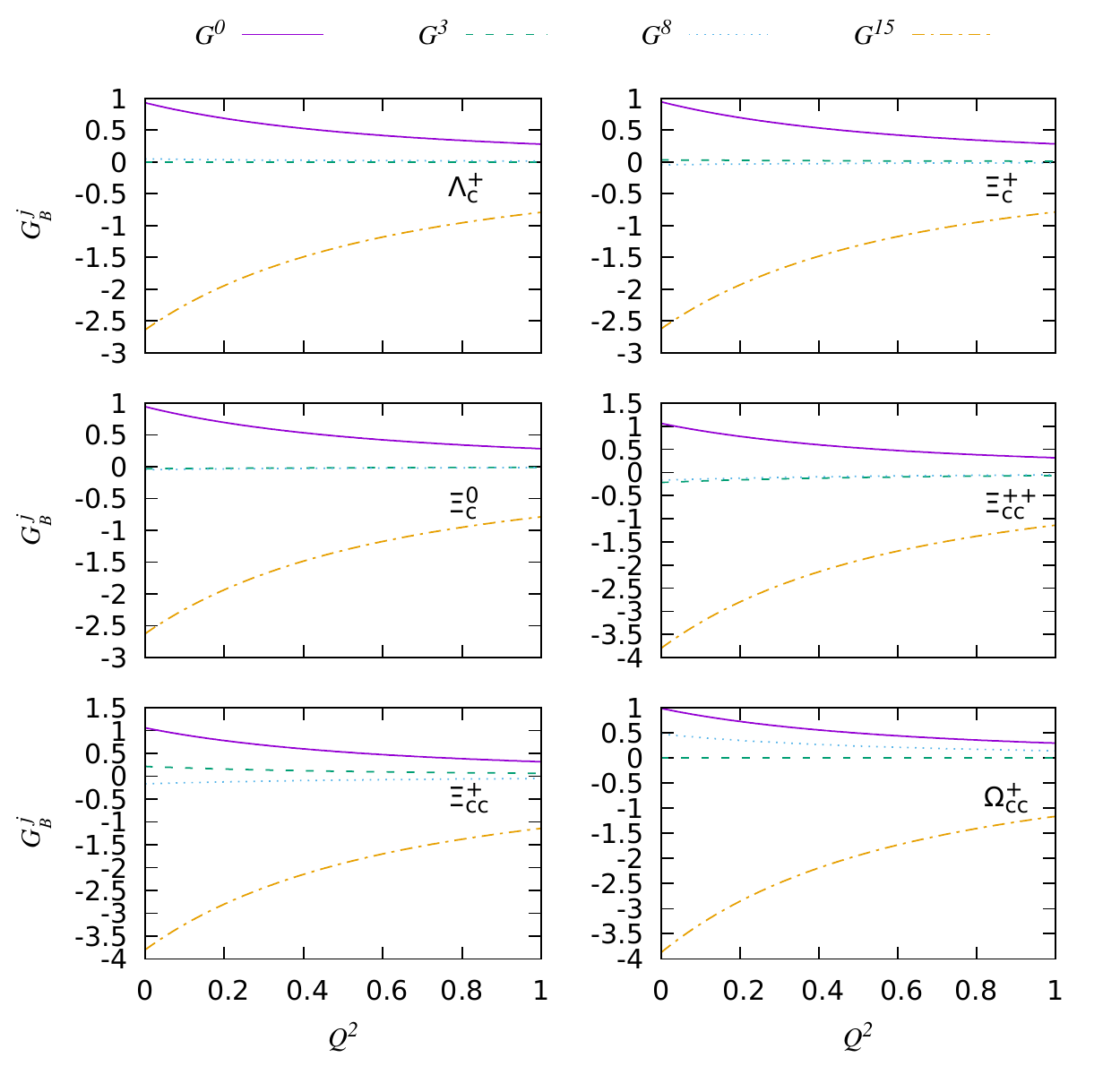}
	\caption{(color online). The axial-vector charges for  the spin $\frac{1}{2}^+$ mixed symmetry 20-plet with the anti-triplet ($\bar{3}$,2) and triplet (3,2) with doubly charmed baryons plotted as function of  $Q^2$ (in units of GeV$^2$).}
	\label{Spin_1by2_Triplet}
\end{figure}

We now discuss the $Q^2$ dependence in the axial-vector charges for each baryon in the range $0\leq Q^2\leq 1$. In Fig. \ref{Spin_1by2_Octet}, we have presented the axial-vector charges for  the spin $\frac{1}{2}^+$ mixed symmetry 20-plet with the light baryon octet (8,0) plotted as function of  $Q^2$. It is clear from Eq. (\ref{ggg}) and the results from Table \ref{QuarkPolar_Spin_1by2_Octet} that because of very small contribution from the charm spin polarization, the singlet and charmed charges are almost same ($g^0\simeq g^{15}$). Both these quantities fall off steadily as the value of $Q^2$ increases. For the triplet and octet charges, since the magnitude is large as compared to that of $g^0$ and $g^{15}$, the decrease/increase is more rapid. Further,  for the members of isospin multiplets, the triplet charges are equal and opposite whereas the octet charges are equal.  In Fig. \ref{Spin_1by2_Sextet}, we have presented the axial-vector charges for  the spin $\frac{1}{2}^+$ mixed symmetry 20-plet with the singly charmed sextet (6,1)  plotted as function of  $Q^2$. 
The singlet and charmed charges in this case have different values because of a charm quark in the constituent structure leading to a significant charm spin polarization. In this case also the triplet charges are equal and opposite whereas the octet charges are equal for the members of isospin multiplets. Depending on the magnitude at $Q^2=0$ all the axial-vector charges decrease with increase in the value of $Q^2$. Higher the magnitude at $Q^2=0$, more rapid is the change. In addition to this the variation of $G_B^j$ with $Q^2$ for all the baryons is more steep for lower value of $Q^2$ ($\simeq 0.6$). For higher values of $Q^2$, $G_B^j$ variation is much less and becomes more or less constant  for values $Q^2=1.$
For the spin $\frac{1}{2}^+$ mixed symmetry 20-plet anti-triplet ($\bar{3}$,2) and triplet (3,2) with doubly charmed baryons presented in Fig. \ref{Spin_1by2_Triplet}, the value of the charmed charge dominates over singlet, triplet and octet charges. The constituent structure includes  two charm quarks and the charm spin polarization is expected to be large. From a cursory look at the plots, one can easily see that the triplet and octet charges do not have much difference in their magnitudes. This can again be explained using the expressions of $g^3$ and $g^8$ from Eq. (\ref{ggg}) in terms of the $u$, $d$ and $s$ spin polarizations. Since there is no contribution of the charm spin polarization in the triplet and octet charges and further the magnitude of the $u$, $d$ and $s$ spin polarizations  is small, they add up having values very close to each other. Only the singlet $g^0$ and charm $g^{15}$ charges have significant values which is mainly because of the large charm polarization in these cases.

\begin{figure}
	\includegraphics {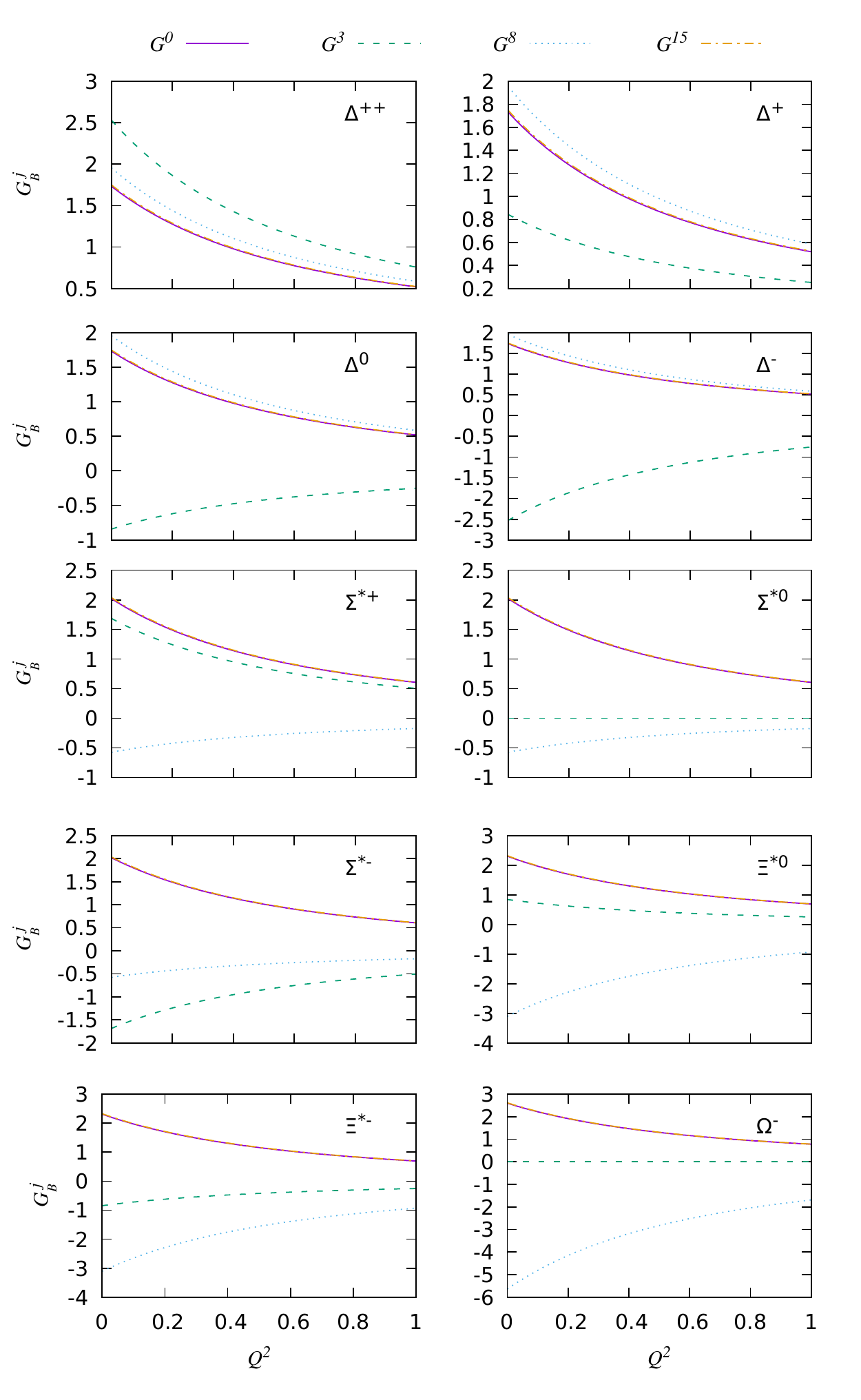}
	\caption{(color online). The axial-vector charges for the spin $\frac{3}{2}^+$  symmetric 20-plet with the light baryon decuplet (10,0) plotted as function of  $Q^2$ (in units of GeV$^2$).}
	\label{Spin_3by2_Decuplet}
\end{figure}

\begin{figure}
	\includegraphics {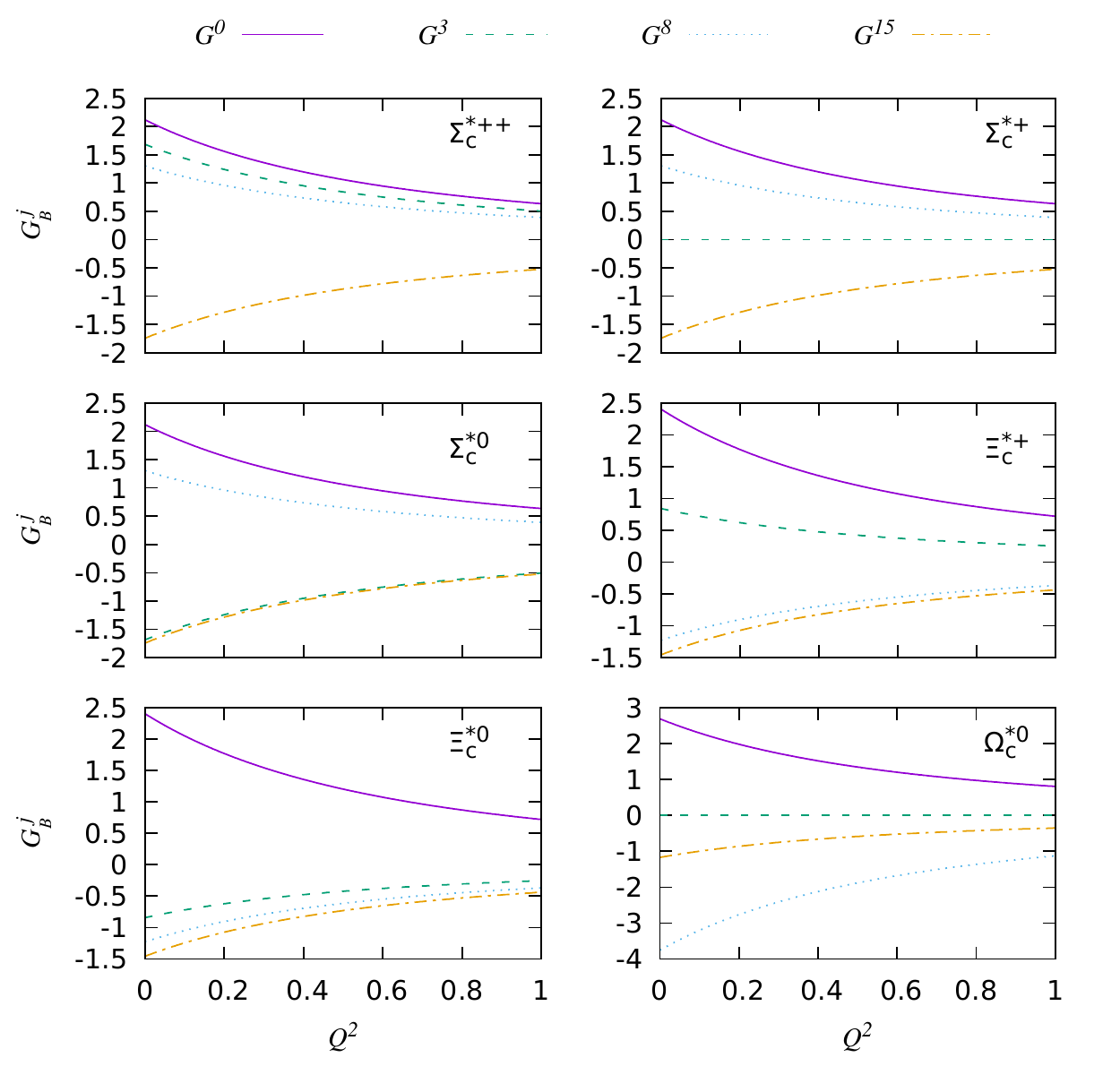}
	\caption{(color online). The axial-vector charges for the spin $\frac{3}{2}^+$  symmetric 20-plet with the singly charmed sextet (6,1) plotted as function of  $Q^2$ (in units of GeV$^2$).}
	\label{Spin_3by2_Sextet}
\end{figure}

\begin{figure}
	\includegraphics {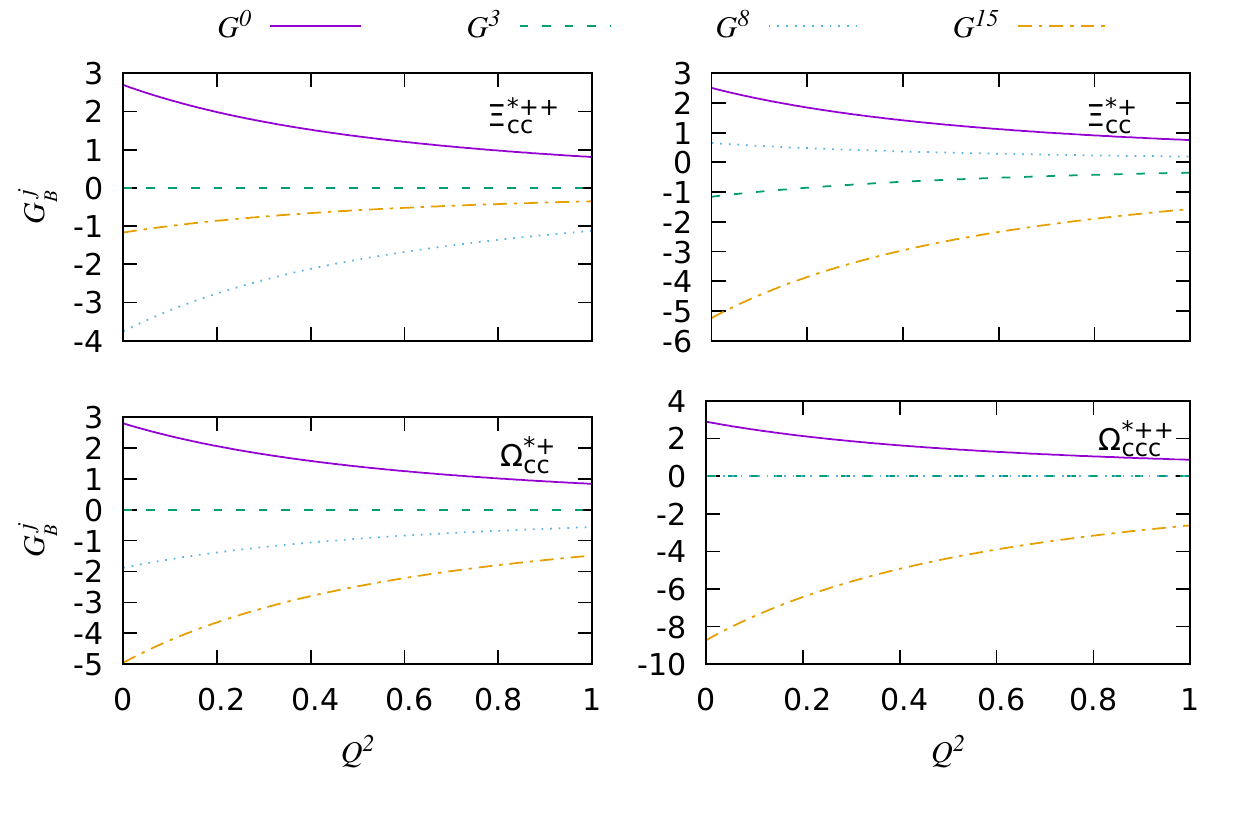}
	\caption{(color online). The axial-vector charges for the spin $\frac{3}{2}^+$  symmetric 20-plet with the doubly charmed triplet (3,2) and triply charmed singlet (1,3) plotted as function of  $Q^2$ (in units of GeV$^2$).}
	\label{Spin_3by2_Triplet_Singlet}
\end{figure}

Further in Figs. \ref{Spin_3by2_Decuplet}, \ref{Spin_3by2_Sextet} and \ref{Spin_3by2_Triplet_Singlet} we have presented the results for the spin $\frac{3}{2}^+$  symmetric 20-plet with the light baryon decuplet (10,0), the singly charmed sextet (6,1) and the doubly charmed triplet (3,2) alongwith the triply charmed singlet (1,3) respectively. The sensitivity to $Q^2$ is in similar lines as that of the spin $\frac{1}{2}^+$ baryons. For the case of the spin $\frac{3}{2}^+$ light baryon decuplet (10,0), small contribution from the charm spin polarization keeps the singlet and charmed charges almost same ($g^0\simeq g^{15}$). The members of isospin multiplets have equal and opposite triplet charges but equal octet charges. 
For the singly charmed sextet (6,1) presented in Fig. \ref{Spin_3by2_Sextet} as a function of  $Q^2$, the singlet charge in all the cases is distinctly large but in the case of triplet, octet and charmed charges the magnitude depends on the number of strange or charm quarks in the constituent structure. In these cases, the quark spin polarizations add or cancel each other depending on the quarks in constituent structure. The degree of variation of the axial-vector charges again depends upon the magnitude at $Q^2=0$. 

\section{Conclusions}

To conclude, our results confirm the presence of intrinsic strange and charm quarks in the light baryons and the other non-constituent quark contributions in the charmed baryons. The understanding of  pivotal role played by SU(3) symmetry breaking will lead to crucial hints in understanding the fundamental structure of the baryons in the nonperturbative QCD regime. The phenomena of chiral symmetry breaking will have far-reaching implications in terms of the elementary degrees of freedom of the composite particles.  Lattice calculations and future experiments  at EIC will not only have the possibility to illuminate the complicated issue of spin structure but also impose significant and decisive restraints in different kinematic regions. 

\section{Data Availability}

This manuscript has no associated data.

\section{Acknowledgement}
 H.D. would like to thank the Science and Engineering Research Board, Department of Science and Technology, Government of India through the grant under MATRICS scheme (Ref No. MTR/2019/000003)  for financial support.

\begin{sidewaystable}
	\scriptsize
	\centering
	
	\begin{center}

		\begin{tabular}{|c|p{5cm}|p{5cm}| p{5cm}|p{4cm}|} 
			\hline
			\hline
			Baryon &\centering {$\Delta u$}& \centering {$\Delta d$}& \centering {$\Delta s$} &  {\qquad \qquad\qquad$\Delta c$ } \\ 
			\hline
			$p(uud)$   & $\cos^2 \phi
			\Big[\frac{4}{3} - \frac{a}{3}(7+ 4\alpha^2+
			\frac{4}{3}\beta^2$  $+ \frac{\zeta^2}{6}+
			\frac{17}{4}\gamma^2) \Big] $
			$+ \sin^2 \phi \Big[\frac{2}{3} - \frac{a}{3}(5 +2 \alpha^2 +
			\frac{2}{3}\beta^2$ $ + \frac{\zeta^2}{12}+
			\frac{17}{8}\gamma^2) \Big]$
			
			& 	$ \cos^2 \phi \Big[-\frac{1}{3} - \frac{a}{3}(2 - \alpha^2 -
			\frac{\beta^2}{3}$ $ - \frac{\zeta^2}{24} - \frac{17}{16}\gamma^2) \Big] $
			$+ \sin^2 \phi \Big[\frac{1}{3} - \frac{a}{3}(4 + \alpha^2 +
			\frac{\beta^2}{3}$ $ + \frac{\zeta^2}{24}+
			\frac{17}{16}\gamma^2) \Big] $ 
			& $ -a \alpha^2 $ & $ -a \gamma^2 $ \\ 
			\hline
			$n(udd)$   			 
			& 	$ \cos^2 \phi \Big[-\frac{1}{3} - \frac{a}{3}(2 - \alpha^2 -
			\frac{\beta^2}{3}$ $ - \frac{\zeta^2}{24} - \frac{17}{16}\gamma^2) \Big] $
			$+ \sin^2 \phi \Big[\frac{1}{3} - \frac{a}{3}(4 + \alpha^2 +
			\frac{\beta^2}{3}$ $ + \frac{\zeta^2}{24}+
			\frac{17}{16}\gamma^2) \Big] $ 
			&$\cos^2 \phi
			\Big[\frac{4}{3} - \frac{a}{3}(7+ 4\alpha^2+
			\frac{4}{3}\beta^2$  $+ \frac{\zeta^2}{6}+
			\frac{17}{4}\gamma^2) \Big] $
			$+ \sin^2 \phi \Big[\frac{2}{3} - \frac{a}{3}(5 +2 \alpha^2 +
			\frac{2}{3}\beta^2$ $ + \frac{\zeta^2}{12}+
			\frac{17}{8}\gamma^2) \Big]$
			& $ -a \alpha^2 $ & $ -a \gamma^2 $ \\ 
			\hline
			$\Sigma^+(uus)$   & $\cos^2 \phi
			\Big[\frac{4}{3} - \frac{a}{3}(8+ 3\alpha^2+
			\frac{4}{3}\beta^2$  $+ \frac{\zeta^2}{6}+
			\frac{17}{4}\gamma^2) \Big] $
			$+ \sin^2 \phi \Big[\frac{2}{3} - \frac{a}{3}(4 +6 \alpha^2 +
			\frac{2}{3}\beta^2$ $ + \frac{\zeta^2}{12}+
			\frac{17}{8}\gamma^2) \Big]$
			
			& 	$ \cos^2 \phi \Big[- \frac{a}{3}(4 - \alpha^2) \Big] $
			$+ \sin^2 \phi \Big[ - \frac{a}{3}(2 + \alpha^2) \Big] $ 
			& 	$ \cos^2 \phi \Big[-\frac{1}{3} - \frac{a}{3}(2\alpha^2 -
			\frac{4}{3}\beta^2$  $ - \frac{\zeta^2}{24} -
			\frac{17}{16}\gamma^2) \Big] $
			$+ \sin^2 \phi \Big[\frac{1}{3} - \frac{a}{3}(4\alpha^2 +
			\frac{4}{3}\beta^2$  $ + \frac{\zeta^2}{24} +
			\frac{17}{16}\gamma^2) \Big] $  & $ -a \gamma^2$ \\ 
			\hline
			
			$\Sigma^0(uds)$  & $\cos^2 \phi
			\Big[\frac{2}{3} - \frac{a}{3}(6+ \alpha^2+
			\frac{2}{3}\beta^2$  $+ \frac{\zeta^2}{12}+
			\frac{17}{8}\gamma^2) \Big] $
			$+ \sin^2 \phi \Big[\frac{1}{3} - \frac{a}{3}(3 + 2 \alpha^2 +
			\frac{\beta^2}{3}$ $ + \frac{\zeta^2}{24}+
			\frac{17}{16}\gamma^2) \Big]$
			
			& 	$ \cos^2 \phi \Big[\frac{2}{3}- \frac{a}{3}(6 + \alpha^2 + \frac{2}{3}\beta^2$  $ + \frac{\zeta^2}{12} - \frac{17}{8}\gamma^2) \Big] $
			$+ \sin^2 \phi \Big[ \frac{1}{3}- \frac{a}{3}(3 + 2\alpha^2 + \frac{\beta^2}{3}$  $ + \frac{\zeta^2}{24} + \frac{17}{16}\gamma^2) \Big] $ 
			& 	$ \cos^2 \phi \Big[-\frac{1}{3} - \frac{a}{3}(2\alpha^2 -
			\frac{4}{3}\beta^2$  $ - \frac{\zeta^2}{24} -
			\frac{17}{16}\gamma^2) \Big] $
			$+ \sin^2 \phi \Big[\frac{1}{3} - \frac{a}{3}(4\alpha^2 +
			\frac{4}{3}\beta^2$  $ + \frac{\zeta^2}{24} +
			\frac{17}{16}\gamma^2) \Big] $ 
			& $ -a \gamma^2 $ \\ 
			
			\hline
			$\Sigma^-(dds)$  			 
			& 	$ \cos^2 \phi \Big[- \frac{a}{3}(4 - \alpha^2) \Big] $
			$+ \sin^2 \phi \Big[ - \frac{a}{3}(2 + \alpha^2) \Big] $ 
			&
			$\cos^2 \phi \Big[\frac{4}{3} - \frac{a}{3}(8+ 3\alpha^2+
			\frac{4}{3}\beta^2$  $+ \frac{\zeta^2}{6}+
			\frac{17}{4}\gamma^2) \Big] $
			$+ \sin^2 \phi \Big[\frac{2}{3} - \frac{a}{3}(4 +6 \alpha^2 +
			\frac{2}{3}\beta^2$ $ + \frac{\zeta^2}{12}+
			\frac{17}{8}\gamma^2) \Big]$
			& 	$ \cos^2 \phi \Big[-\frac{1}{3} - \frac{a}{3}(2\alpha^2 -
			\frac{4}{3}\beta^2$  $ - \frac{\zeta^2}{24} -
			\frac{17}{16}\gamma^2) \Big] $
			$+ \sin^2 \phi \Big[\frac{1}{3} - \frac{a}{3}(4\alpha^2 +
			\frac{4}{3}\beta^2$  $ + \frac{\zeta^2}{24} +
			\frac{17}{16}\gamma^2) \Big] $  & $ -a \gamma^2 $ \\ 
			\hline
			$\Xi^0(uss)$   & $\cos^2 \phi
			\Big[-\frac{1}{3} - \frac{a}{3}(-2+ 3\alpha^2 -
			\frac{\beta^2}{3}$  $ - \frac{\zeta^2}{24} -
			\frac{17}{16}\gamma^2) \Big] $
			$+ \sin^2 \phi \Big[\frac{1}{3} - \frac{a}{3}(2 + 3 \alpha^2 +
			\frac{\beta^2}{3}$ $ + \frac{\zeta^2}{24}+
			\frac{17}{16}\gamma^2) \Big]$
			
			& 	$ \cos^2 \phi \Big[- \frac{a}{3}(-1 + 4\alpha^2) \Big] $
			$+ \sin^2 \phi \Big[ - \frac{a}{3}(1 + 2 \alpha^2) \Big] $ 
			& 	$ \cos^2 \phi \Big[\frac{4}{3} - \frac{a}{3}(7\alpha^2 +
			\frac{16}{3}\beta^2$  $ + \frac{\zeta^2}{6} +
			\frac{17}{4}\gamma^2) \Big] $
			$+ \sin^2 \phi \Big[\frac{2}{3} - \frac{a}{3}(5\alpha^2 +
			\frac{8}{3}\beta^2$  $ + \frac{\zeta^2}{12} +
			\frac{17}{8}\gamma^2) \Big] $  & $ -a \gamma^2$ \\ 
			\hline
			$\Xi^-(dss)$   
			& 	$ \cos^2 \phi \Big[- \frac{a}{3}(-1 + 4\alpha^2) \Big] $
			$+ \sin^2 \phi \Big[ - \frac{a}{3}(1 + 2 \alpha^2) \Big] $ 
			& $\cos^2 \phi
			\Big[-\frac{1}{3} - \frac{a}{3}(-2+ 3\alpha^2 -
			\frac{\beta^2}{3}$  $ - \frac{\zeta^2}{24} -
			\frac{17}{16}\gamma^2) \Big] $
			$+ \sin^2 \phi \Big[\frac{1}{3} - \frac{a}{3}(2 + 3 \alpha^2 +
			\frac{\beta^2}{3}$ $ + \frac{\zeta^2}{24}+
			\frac{17}{16}\gamma^2) \Big]$
			& 	$ \cos^2 \phi \Big[\frac{4}{3} - \frac{a}{3}(7\alpha^2 +
			\frac{16}{3}\beta^2$  $ + \frac{\zeta^2}{6} +
			\frac{17}{4}\gamma^2) \Big] $
			$+ \sin^2 \phi \Big[\frac{2}{3} - \frac{a}{3}(5\alpha^2 +
			\frac{8}{3}\beta^2$  $ + \frac{\zeta^2}{12} +
			\frac{17}{8}\gamma^2) \Big] $  & $ -a \gamma^2 $ \\ 
			\hline
			
			$\Lambda(uds)$  
			& 	$ \cos^2 \phi \Big[- a \alpha^2\Big] $
			$+ \sin^2 \phi \Big[ \frac{1}{3}- \frac{a}{3}(2 + 2 \alpha^2 +\frac{\beta^2}{3} +\frac{\zeta^2}{24}+\frac{17}{16} \gamma^2) \Big] $ 
			& $\cos^2 \phi	\Big[ - a \alpha^2 \Big] $
			$+ \sin^2 \phi \Big[\frac{1}{3} - \frac{a}{3}(2 + 2 \alpha^2 +
			\frac{\beta^2}{3}$ $ + \frac{\zeta^2}{24}+
			\frac{17}{16}\gamma^2) \Big]$
			& 	$ \cos^2 \phi \Big[1 - a(2\alpha^2 +
			\frac{4}{3}\beta^2$  $ + \frac{\zeta^2}{24} +
			\frac{17}{16}\gamma^2) \Big] $
			$+ \sin^2 \phi \Big[\frac{1}{3} - \frac{a}{3}(4\alpha^2 +
			\frac{4}{3}\beta^2$  $ + \frac{\zeta^2}{24} +
			\frac{17}{16}\gamma^2) \Big] $  & $ -a \gamma^2 $ \\ 
			\hline
			
		\end{tabular}
		\caption{ \scriptsize The quark spin polarization for spin $\frac{1}{2}^{+}$ octet baryons (8,0) within  $SU(4)$ representation in the chiral quark constituent model.}	\label{table_baryon_octet_spin_half}
		
	\end{center}

\end{sidewaystable}


\begin{sidewaystable}
	\scriptsize
	\centering
	
	\begin{center}
		\begin{tabular}{|c|p{5cm}|p{5cm}| p{5cm}|p{4cm}|} 
			\hline
			\hline
			Baryon &  \centering {$\Delta u$}& \centering {$\Delta d$}& \centering {$\Delta s$} &  {\qquad \qquad\qquad$\Delta c$ }
			\\	\hline  
			$\Sigma_c^{++}(uuc)$   & $\cos^2 \phi
			\Big[\frac{4}{3} - \frac{a}{3}(8+ 4\alpha^2+
			\frac{4}{3}\beta^2$  $+ \frac{\zeta^2}{6}+
			\frac{13}{4}\gamma^2) \Big] $
			$+ \sin^2 \phi \Big[\frac{2}{3} - \frac{a}{3}(4 +2 \alpha^2 +
			\frac{2}{3}\beta^2$ $ + \frac{\zeta^2}{12}+
			\frac{25}{8}\gamma^2) \Big]$
			& 	$ \cos^2 \phi \Big[- \frac{a}{3}(4 - \gamma^2) \Big] $
			$+ \sin^2 \phi \Big[ - \frac{a}{3}(2 + \gamma^2) \Big] $ 
			&
			$ \cos^2 \phi \Big[- \frac{a}{3}(4 \alpha^2 - \gamma^2) \Big] $
			$+ \sin^2 \phi \Big[ - \frac{a}{3}(2 \alpha^2 + \gamma^2) \Big] $
			& 	$ \cos^2 \phi \Big[-\frac{1}{3} + \frac{a}{24}(3\zeta^2 
			+\gamma^2) \Big] $
			$+ \sin^2 \phi \Big[\frac{1}{3} - \frac{a}{24}(3\zeta^2 
			+ 49 \gamma^2)  \Big] $ 
			\\ 
			\hline
			
			$\Sigma_c^{+}(udc)$   & $\cos^2 \phi
			\Big[\frac{2}{3} - \frac{a}{3}(6+ 2\alpha^2+
			\frac{2}{3}\beta^2$  $+ \frac{\zeta^2}{12}+
			\frac{9}{8}\gamma^2) \Big] $
			$+ \sin^2 \phi \Big[\frac{1}{3} - \frac{a}{3}(3 + \alpha^2 +
			\frac{\beta^2}{3}$ $ + \frac{\zeta^2}{24}+
			\frac{33}{16}\gamma^2) \Big]$
			& 	$ \cos^2 \phi \Big[\frac{2}{3}- \frac{a}{3}(6 + 2\alpha^2 + \frac{2}{3}\beta^2$  $ + \frac{\zeta^2}{12} + \frac{9}{8}\gamma^2) \Big] $
			$+ \sin^2 \phi \Big[ \frac{1}{3}- \frac{a}{3}(3 + \alpha^2 + \frac{\beta^2}{3}$  $ + \frac{\zeta^2}{24} + \frac{33}{16}\gamma^2) \Big] $ 
			& 	$ \cos^2 \phi \Big[ - \frac{a}{3}(4\alpha^2 -
			\gamma^2) \Big] $
			$+ \sin^2 \phi \Big[ - \frac{a}{3}(2\alpha^2 +
			\gamma^2) \Big] $ 
			& 
			$ \cos^2 \phi \Big[-\frac{1}{3} + \frac{a}{24}(3\zeta^2   +
			\gamma^2) \Big] $
			$+ \sin^2 \phi \Big[\frac{1}{3} - \frac{a}{24}(3\zeta^2   +
			49 \gamma^2) \Big] $
			\\ 
			\hline
			$\Sigma_c^{0}(ddc)$   	
			& 	$ \cos^2 \phi \Big[- \frac{a}{3}(4 - \gamma^2) \Big] $
			$+ \sin^2 \phi \Big[ - \frac{a}{3}(2 + \gamma^2) \Big] $		 
			&
			$\cos^2 \phi
			\Big[\frac{4}{3} - \frac{a}{3}(8+ 4\alpha^2+
			\frac{4}{3}\beta^2$  $+ \frac{\zeta^2}{6}+
			\frac{13}{4}\gamma^2) \Big] $
			$+ \sin^2 \phi \Big[\frac{2}{3} - \frac{a}{3}(4 +2 \alpha^2 +
			\frac{2}{3}\beta^2$ $ + \frac{\zeta^2}{12}+
			\frac{25}{8}\gamma^2) \Big]$
			&
			$ \cos^2 \phi \Big[- \frac{a}{3}(4 \alpha^2 - \gamma^2) \Big] $
			$+ \sin^2 \phi \Big[ - \frac{a}{3}(2 \alpha^2 + \gamma^2) \Big] $
			& 	$ \cos^2 \phi \Big[-\frac{1}{3} + \frac{a}{24}(3\zeta^2 
			+\gamma^2) \Big] $
			$+ \sin^2 \phi \Big[\frac{1}{3} - \frac{a}{24}(3\zeta^2 
			+ 49 \gamma^2)  \Big] $ 
			\\ 
			\hline
			$\Xi_c^{\prime +}(cus)$   & $\cos^2 \phi
			\Big[\frac{2}{3} - \frac{a}{3}(4+ 4\alpha^2 -
			\frac{2}{3}\beta^2$  $ + \frac{\zeta^2}{12} -
			\frac{9}{8}\gamma^2) \Big] $
			$+ \sin^2 \phi \Big[\frac{1}{3} - \frac{a}{3}(2 + 2\alpha^2 +
			\frac{\beta^2}{3}$ $ + \frac{\zeta^2}{24}+
			\frac{33}{16}\gamma^2) \Big]$
			
			& 	$ \cos^2 \phi \Big[- \frac{a}{3}(2 + 2\alpha^2 -\gamma^2) \Big] $
			$+ \sin^2 \phi \Big[ - \frac{a}{3}(1 +  \alpha^2 + \gamma^2) \Big] $ 
			& 	$ \cos^2 \phi \Big[\frac{2}{3} - \frac{a}{3}(6\alpha^2 +
			\frac{8}{3}\beta^2$  $ + \frac{\zeta^2}{12} +
			\frac{9}{8}\gamma^2) \Big] $
			$+ \sin^2 \phi \Big[\frac{1}{3} - \frac{a}{3}(3\alpha^2 +
			\frac{4}{3}\beta^2$  $ + \frac{\zeta^2}{24} +
			\frac{33}{16}\gamma^2) \Big] $ 
			& 
			$ \cos^2 \phi \Big[-\frac{1}{3} + \frac{a}{24}(3\zeta^2 +
			\gamma^2) \Big] $
			$+ \sin^2 \phi \Big[\frac{1}{3} - \frac{a}{24}(3\zeta^2 +
			49\gamma^2) \Big] $ 
			\\ 
			\hline
			$\Xi_c^{\prime 0}(cds)$   
			& 	$ \cos^2 \phi \Big[- \frac{a}{3}(2 + 2\alpha^2 -\gamma^2) \Big] $
			$+ \sin^2 \phi \Big[ - \frac{a}{3}(1 +  \alpha^2 + \gamma^2) \Big] $
			&
			$\cos^2 \phi \Big[\frac{2}{3} - \frac{a}{3}(4+ 4\alpha^2 -
			\frac{2}{3}\beta^2$  $ + \frac{\zeta^2}{12} -
			\frac{9}{8}\gamma^2) \Big] $
			$+ \sin^2 \phi \Big[\frac{1}{3} - \frac{a}{3}(2 + 2\alpha^2 +
			\frac{\beta^2}{3}$ $ + \frac{\zeta^2}{24}+
			\frac{33}{16}\gamma^2) \Big]$
			
			& 	$ \cos^2 \phi \Big[\frac{2}{3} - \frac{a}{3}(6\alpha^2 +
			\frac{8}{3}\beta^2$  $ + \frac{\zeta^2}{12} +
			\frac{9}{8}\gamma^2) \Big] $
			$+ \sin^2 \phi \Big[\frac{1}{3} - \frac{a}{3}(3\alpha^2 +
			\frac{4}{3}\beta^2$  $ + \frac{\zeta^2}{24} +
			\frac{33}{16}\gamma^2) \Big] $ 
			& 
			$ \cos^2 \phi \Big[-\frac{1}{3} + \frac{a}{24}(3\zeta^2 +
			\gamma^2) \Big] $
			$+ \sin^2 \phi \Big[\frac{1}{3} - \frac{a}{24}(3\zeta^2 +
			49\gamma^2) \Big] $ 
			\\ 
			\hline
			
			$\Omega_c^0(css)$  
			& 	$ \cos^2 \phi \Big[- \frac{a}{3}(4\alpha^2 -\gamma^2)\Big] $
			$+ \sin^2 \phi \Big[ - \frac{a}{3} (2\alpha^2 +\gamma^2) \Big] $ 
			& 
			$ \cos^2 \phi \Big[- \frac{a}{3}(4\alpha^2 -\gamma^2)\Big] $
			$+ \sin^2 \phi \Big[ - \frac{a}{3} (2\alpha^2 +\gamma^2) \Big] $ 
			& 	$ \cos^2 \phi \Big[\frac{4}{3} - \frac{a}{3}(8\alpha^2 +
			\frac{16}{3}\beta^2$  $ + \frac{\zeta^2}{6} +
			\frac{13}{4}\gamma^2) \Big] $
			$+ \sin^2 \phi \Big[\frac{2}{3} - \frac{a}{3}(4\alpha^2 +
			\frac{8}{3}\beta^2$  $ + \frac{\zeta^2}{12} +
			\frac{33}{8}\gamma^2) \Big] $  & 
			$ \cos^2 \phi \Big[-\frac{1}{3} + \frac{a}{24}(3\zeta^2   +
			\gamma^2) \Big] $
			$+ \sin^2 \phi \Big[\frac{1}{3} - \frac{a}{24}(3\zeta^2   +
			49 \gamma^2) \Big] $
			
			\\ 
			\hline
		\end{tabular}
		\caption{\scriptsize The quark spin polarization for singly charmed spin $\frac{1}{2}^{+}$  baryons (6,1) within  $SU(4)$ representation in the chiral quark constituent model}\label{table_single_charmed_spin_half}
	\end{center}
	
\end{sidewaystable}



\begin{sidewaystable}
	\scriptsize
	\centering
	
	\begin{center}
		\begin{tabular}{|c |p{5cm}|p{5cm}| p{5cm}|p{4cm}|} 
			\hline
			\hline
			Baryon &  \centering {$\Delta u$}& \centering {$\Delta d$}& \centering {$\Delta s$} &  {\qquad \qquad\qquad$\Delta c$ } \\
			\hline
			
			$\Lambda_c^{+}(udc)$   & $\cos^2 \phi
			\Big [-a\gamma^2 \Big] $
			$+ \sin^2 \phi \Big[\frac{1}{3} - \frac{a}{3}(3 + \alpha^2 +
			\frac{\beta^2}{3}$ $ + \frac{\zeta^2}{24}+
			\frac{33}{16}\gamma^2) \Big]$
			
			& 	$ \cos^2 \phi \Big[  - a\gamma^2 \Big] $
			$+ \sin^2 \phi \Big[  \frac{1}{3}- \frac{a}{3}(3 + \alpha^2 +
			\frac{\beta^2}{3}$ $ + \frac{\zeta^2}{24}+
			\frac{33}{16}\gamma^2) \Big] $ 
			& 	$ \cos^2 \phi \Big[  - a\gamma^2  \Big] $
			$+ \sin^2 \phi \Big[  - \frac{a}{3}(2\alpha^2 +
			\gamma^2) \Big] $  &
			$ \cos^2 \phi \Big[ 1-  \frac{3}{8}a(\zeta^2 +
			11\gamma^2) \Big] $
			$+ \sin^2 \phi \Big[\frac{1}{3} - \frac{a}{24}(3\zeta^2 +
			49 \gamma^2) \Big] $  
			
			\\ 
			\hline
			
			$\Xi_c^{+}(usc)$   &  
			$\cos^2 \phi \Big [-a\gamma^2 \Big] $
			$+ \sin^2 \phi \Big[\frac{1}{3} - \frac{a}{3}(2 +2 \alpha^2 +
			\frac{\beta^2}{3}$ $ + \frac{\zeta^2}{24}+
			\frac{33}{16}\gamma^2) \Big]$
			
			& 	$ \cos^2 \phi \Big[ - a\gamma^2 \Big] $
			$+ \sin^2 \phi \Big[  \frac{1}{3}- \frac{a}{3}(1 + \alpha^2 +
			\gamma^2) \Big] $ 
			& 	$ \cos^2 \phi \Big[  - a\gamma^2  \Big] $
			$+ \sin^2 \phi \Big[  - \frac{a}{3}(3\alpha^2 + \frac{4}{3}\beta^2 + \frac{\zeta^2}{24}+
			\frac{33}{16} \gamma^2) \Big] $  &
			$ \cos^2 \phi \Big[ 1-  \frac{3}{8}a(\zeta^2 +
			11\gamma^2) \Big] $
			$+ \sin^2 \phi \Big[\frac{1}{3} - \frac{a}{24}(3\zeta^2 +
			49 \gamma^2) \Big] $ 
			
			\\ 
			
			\hline
			$\Xi_c^{0}(dsc)$   			 
			& 
			$ \cos^2 \phi \Big[ - a\gamma^2 \Big] $
			$+ \sin^2 \phi \Big[  \frac{1}{3}- \frac{a}{3}(1 + \alpha^2 +
			\gamma^2) \Big] $ 	
			&
			$\cos^2 \phi \Big [-a\gamma^2 \Big] $
			$+ \sin^2 \phi \Big[\frac{1}{3} - \frac{a}{3}(2 +2 \alpha^2 +
			\frac{\beta^2}{3}$ $ + \frac{\zeta^2}{24}+
			\frac{33}{16}\gamma^2) \Big]$
			
			& 	$ \cos^2 \phi \Big[  - a\gamma^2  \Big] $
			$+ \sin^2 \phi \Big[  - \frac{a}{3}(3\alpha^2 + \frac{4}{3}\beta^2 + \frac{\zeta^2}{24}+
			\frac{33}{16} \gamma^2) \Big] $  &
			$ \cos^2 \phi \Big[ 1-  \frac{3}{8}a(\zeta^2 +
			11\gamma^2) \Big] $
			$+ \sin^2 \phi \Big[\frac{1}{3} - \frac{a}{24}(3\zeta^2 +
			49 \gamma^2) \Big] $ 
			
			\\ 
			\hline
			$\Xi_{cc}^{++}(ucc)$   &
			$\cos^2 \phi
			\Big[-\frac{1}{3} + \frac{a}{3}(2+ \alpha^2 -
			\frac{\beta^2}{3}$  $ + \frac{\zeta^2}{24} -
			\frac{47}{16}\gamma^2) \Big] $
			$+ \sin^2 \phi \Big[\frac{1}{3} - \frac{a}{3}(2 +  \alpha^2 +
			\frac{\beta^2}{3}$ $ + \frac{\zeta^2}{24}+
			\frac{49}{16}\gamma^2) \Big]$
			
			& 	$ \cos^2 \phi \Big[\frac{a}{3}(1 - 4\gamma^2) \Big] $
			$+ \sin^2 \phi \Big[ - \frac{a}{3}(1 + 2 \gamma^2) \Big] $ 
			& 	$ \cos^2 \phi \Big[   \frac{a}{3}(\alpha^2 - 4\gamma^2) \Big] $
			$+ \sin^2 \phi \Big[  - \frac{a}{3}(\alpha^2 +
			2 \gamma^2) \Big] $  &  
			
			$\cos^2 \phi
			\Big[\frac{4}{3} - \frac{a}{6}(3\zeta^2 +
			31\gamma^2) \Big] $
			$+ \sin^2 \phi \Big[\frac{2}{3} - \frac{a}{12}(3\zeta^2 +
			37 \gamma^2) \Big]$
			\\ 
			\hline
			$\Xi_{cc}^{+}(dcc)$  
			
			&
			$ \cos^2 \phi \Big[\frac{a}{3}(1 - 4\gamma^2) \Big] $
			$+ \sin^2 \phi \Big[ - \frac{a}{3}(1 + 2 \gamma^2) \Big] $ 
			&  
			$\cos^2 \phi
			\Big[-\frac{1}{3} + \frac{a}{3}(2+ \alpha^2 -
			\frac{\beta^2}{3}$  $ + \frac{\zeta^2}{24} -
			\frac{47}{16}\gamma^2) \Big] $
			$+ \sin^2 \phi \Big[\frac{1}{3} - \frac{a}{3}(2 +  \alpha^2 +
			\frac{\beta^2}{3}$ $ + \frac{\zeta^2}{24}+
			\frac{49}{16}\gamma^2) \Big]$
			
			& 	$ \cos^2 \phi \Big[   \frac{a}{3}(\alpha^2 - 4\gamma^2) \Big] $
			$+ \sin^2 \phi \Big[  - \frac{a}{3}(\alpha^2 +
			2 \gamma^2) \Big] $  &  
			
			$\cos^2 \phi
			\Big[\frac{4}{3} - \frac{a}{6}(3\zeta^2 +
			31\gamma^2) \Big] $
			$+ \sin^2 \phi \Big[\frac{2}{3} - \frac{a}{12}(3\zeta^2 +
			37 \gamma^2) \Big]$

			\\ 
			\hline
			$\Omega_{cc}^+(scc)$  
			&  
			$ \cos^2 \phi \Big[\frac{a}{3}(\alpha^2 -4\gamma^2)\Big] $
			$+ \sin^2 \phi \Big[ - \frac{a}{3} (\alpha^2 + 2\gamma^2) \Big] $ 
			& 
			$ \cos^2 \phi \Big[ \frac{a}{3}(\alpha^2 - 4\gamma^2)\Big] $
			$+ \sin^2 \phi \Big[ - \frac{a}{3} (\alpha^2 + 2\gamma^2) \Big] $ 
			& 	$ \cos^2 \phi \Big[-\frac{1}{3} + \frac{a}{3}(2\alpha^2 +
			\frac{4}{3}\beta^2$  $ + \frac{\zeta^2}{24} -
			\frac{47}{16}\gamma^2) \Big] $
			$+ \sin^2 \phi \Big[\frac{1}{3} - \frac{a}{3}(2\alpha^2 +
			\frac{4}{3}\beta^2$  $ + \frac{\zeta^2}{24} +
			\frac{47}{16}\gamma^2) \Big] $  & 
			$ \cos^2 \phi \Big[\frac{4}{3} - \frac{a}{2}(\zeta^2   +
			\frac{31}{3} \gamma^2) \Big] $
			$+ \sin^2 \phi \Big[\frac{2}{3} - \frac{a}{4}(\zeta^2   +
			\frac{37}{3} \gamma^2) \Big] $
			\\ 
			\hline
			
		\end{tabular}
		\caption{ \scriptsize The quark spin polarization for double charmed spin $\frac{1}{2}^{+}$ baryons (3,2) within $SU(4)$ representation in the chiral quark constituent model.}\label{table_double_charmed_spin_half}
	\end{center}

\end{sidewaystable}

\begin{sidewaystable}
	\scriptsize
	\centering
	\begin{center}
		\begin{tabular} 	{|c |p{5cm}|p{5cm}| p{5cm}|p{4cm}|} 
			\hline
			\hline
			Baryon &  \centering {$\Delta u$}& \centering {$\Delta d$}& \centering {$\Delta s$} &  {\qquad \qquad\qquad$\Delta c$ } \\
			\hline
			$\Delta^{++}(uuu)$ 
			&$3 - a\left(6+ 3\alpha^2+ \beta^2+ \frac{\zeta^2}{8}+ \frac{51}{16}\gamma^2\right)$ 
			& $- 3a$ 
			& $-3a\alpha^2 $ 
			& $-3a \gamma^2$ 
			\\\hline
			$\Delta^{+}(uud)$ 
			& $ 2 -a \left(5 + 2\alpha^2+ \frac{2}{3}\beta^2+\frac{\zeta^2}{12}+ \frac{17}{8}\gamma^2\right)$
			& $1 - a\left(4+\alpha^2+ \frac{\beta^2}{3}+ \frac{\zeta^2}{24}+\frac{17}{16}\gamma^2\right)$ 
			& $- 3a\alpha^2$  
			& $- 3a\gamma^2$
			\\\hline
			$\Delta^{0}(udd)$ 
			& $1 - a\left(4+\alpha^2+ \frac{\beta^2}{3}+ \frac{\zeta^2}{24}+\frac{17}{16}\gamma^2\right)$ 
			&$ 2 -a \left(5 + 2\alpha^2+ \frac{2}{3}\beta^2+\frac{\zeta^2}{12}+ \frac{17}{8}\gamma^2\right)$ 
			&$-3a\alpha^2$
			&$-3a\gamma^2$
			\\ 
			\hline
			$\Delta^{-}(ddd)$
			& $- 3a$ 
			&$3 - a\left(6+ 3\alpha^2+ \beta^2+ \frac{\zeta^2}{8}+ \frac{51}{16}\gamma^2\right)$ 
			& $-3a\alpha^2 $ 
			& $-3a \gamma^2$ 
			\\\hline
			
			$\Sigma^{*+}(uus)$ 
			& $ 2 - a \left(4+ 3\alpha^2+ \frac{2}{3}\beta^2+\frac{\zeta^2}{12}+ \frac{17}{8}\gamma^2\right)$  
			& $- a\left(2 + \alpha^2 \right) $ 
			& $1- a\left(4\alpha^2+ \frac{4}{3}\beta^2+ \frac{\zeta^2}{24}+ \frac{17}{16}\gamma^2\right)$ 
			&  $ -3a\gamma^2$
			\\\hline

			$\Sigma^{*0}(uds)$ 
			&$1-a\left(3+2\alpha^2+\frac{\beta^2}{3}+ \frac{\zeta^2}{24}+\frac{17}{16}\gamma^2\right)$ 
			& $1- a\left(3+ 2\alpha^2+\frac{\beta^2}{3}+ \frac{\zeta^2}{24}+ \frac{17}{16}\gamma^2\right)$  
			& $1- a\left(4\alpha^2+ \frac{4}{3}\beta^2+\frac{\zeta^2}{24}+ \frac{17}{16}\gamma^2\right)$ 
			& $-3a\gamma^2$
			\\\hline
			
			$\Sigma^{*-}(dds)$
			& $- a\left(2 + \alpha^2 \right) $ 
			& $ 2 - a \left(4+ 3\alpha^2+ \frac{2}{3}\beta^2+\frac{\zeta^2}{12}+ \frac{17}{8}\gamma^2\right)$ 
			& $1- a\left(4\alpha^2+ \frac{4}{3}\beta^2+ \frac{\zeta^2}{24}+ \frac{17}{16}\gamma^2\right)$ 
			&  $ -3a\gamma^2$
			\\\hline

			$\Xi^{*0}(uss)$ 
			& $1-a \left(2+ 3\alpha^2+ \frac{\beta^2}{3}+\frac{\zeta^2}{24}+ \frac{17}{16}\gamma^2\right)$ 
			& $- a\left(1+2\alpha^2\right)$ 
			& $2- a\left(5\alpha^2+ \frac{8}{3}\beta^2+\frac{\zeta^2}{12}+ \frac{17}{8}\gamma^2\right)$ 
			& $-3a\gamma^2$ 
			\\\hline

			$\Xi^{*-}(dss)$ 
			& $- a\left(1+2\alpha^2\right)$ 
			& $1-a \left(2+ 3\alpha^2+ \frac{\beta^2}{3}+\frac{\zeta^2}{24}+ \frac{17}{16}\gamma^2\right)$
			& $2- a\left(5\alpha^2+ \frac{8}{3}\beta^2+\frac{\zeta^2}{12}+ \frac{17}{8}\gamma^2\right)$ 
			& $-3a\gamma^2$
			\\\hline	
			
			$\Omega^{*-}(sss)$ 
			& $-3a\alpha^2$ 
			& $-3a\alpha^2$ 
			& $3-a\left(6\alpha^2+ 4\beta^2+ \frac{\zeta^2}{8} +\frac{51}{16}\gamma^2\right)$ 
			& $- 3a\gamma^2$\\
			\hline
			
		\end{tabular}
		\caption{\scriptsize The quark spin polarization for spin $\frac{3}{2}^{+}$ baryons, (10,0), within $SU(4)$ representation in the chiral quark constituent model.}\label{table_decuplet_3_2}
	\end{center}

\end{sidewaystable}

\begin{sidewaystable}
	\scriptsize
	\centering
	\begin{center}
		\begin{tabular} 	{|c |p{5cm}|p{5cm}| p{5cm}|p{4cm}|} 
			\hline
			\hline
			Baryon &  \centering {$\Delta u$}& \centering {$\Delta d$}& \centering {$\Delta s$} &  {\qquad \qquad\qquad$\Delta c$ } \\
			\hline        
			$\Sigma^{*++}_{c}(uuc)$ 
			& $2 -a\left(4+ 2\alpha^2+\frac{2}{3}\beta^2+ \frac{\zeta^2}{12}+ \frac{25}{8}\gamma^2\right)$ 
			& $- a\left(2+ \gamma^2 \right)$ 
			& $- a\left(2\alpha^2+ \gamma^2 \right)$ 
			& $1 - \frac{a}{8}\left(3\zeta^2+ 49\gamma^2 \right)$
			\\\hline

			$\Sigma^{*+}_{c}(udc)$ 
			& $1+ -a\left(3+ \alpha^2+ \frac{\beta^2}{3}+ \frac{\zeta^2}{24} +             \frac{33}{16}\gamma^2\right)$ 
			&  $1- a\left(3+ \alpha^2+\frac{\beta^2}{3}+ \frac{\zeta^2}{24}+ \frac{33}{16}\gamma^2\right)$ 
			& $- a\left(2 \alpha^2+ \gamma^2\right)$ 
			&  $1- \frac{a}{8}\left(3\zeta^2+ 49\gamma^2\right)$
			\\\hline
			
			$\Sigma^{*0}_{c}(ddc)$ 
			& $- a\left(2+ \gamma^2 \right)$ 
			& $2 -a\left(4+ 2\alpha^2+\frac{2}{3}\beta^2+ \frac{\zeta^2}{12}+ \frac{25}{8}\gamma^2\right)$
			& $- a\left(2\alpha^2+ \gamma^2 \right)$ 
			& $1 - \frac{a}{8}\left(3\zeta^2+ 49\gamma^2 \right)$
			\\\hline
			
			$\Xi^{*+}_{c}(usc)$ 
			& $1-a\left(2+ 2\alpha^2+ \frac{\beta^2}{3}+ \frac{\zeta^2}{24} +\frac{33}{16}\gamma^2\right)$ 
			& $- a\left(1+ \alpha^2 +\gamma^2\right) $ 
			& $1- a\left (3\alpha^2+ \frac{4}{3}\beta^2+\frac{\zeta^2}{24}+ \frac{33}{16}\gamma^2\right)$
			& $1-\frac{a}{8}\left(3\zeta^2+ 49\gamma^2\right)$\\
			\hline
			
			$\Xi^{*0}_{c}(cds)$ 
			&$- a\left(1+ \alpha^2 +\gamma^2\right) $
			&$1-a\left(2+ 2\alpha^2+ \frac{\beta^2}{3}+ \frac{\zeta^2}{24} +\frac{33}{16}\gamma^2\right)$
			& $1- a\left (3\alpha^2+ \frac{4}{3}\beta^2+\frac{\zeta^2}{24}+ \frac{33}{16}\gamma^2\right)$
			& $1-\frac{a}{8}\left(3\zeta^2+ 49\gamma^2\right)$
			
			\\\hline

			$\Omega^{*0}_{c} (css)$ 
			&$- a(2\alpha^2+ \gamma^2)$ 
			&$ - a(2\alpha^2+ \gamma^2)$ 
			&$2 - a(4\alpha^2+\frac{8}{3}\beta^2+ \frac{\zeta^2}{12}+\frac{25}{8}\gamma^2)$ 
			& $1  - \frac{a}{8}\left(3\zeta^2+ 49\gamma^2\right)$
			\\\hline 
			
		\end{tabular}
		\caption{\scriptsize The quark spin polarization for singly charmed spin $\frac{3}{2}^{+}$ baryons, (6,1), within $SU(4)$ representation in the chiral quark constituent model.}\label{table_single_charm_3_2}
	\end{center}

\end{sidewaystable}

\begin{sidewaystable}
	\scriptsize
	\centering
	\begin{center}
		\begin{tabular} 	{|c |p{5cm}|p{5cm}| p{5cm}|p{4cm}|} 
			\hline
			\hline
			Baryon &  \centering {$\Delta u$}& \centering {$\Delta d$}& \centering {$\Delta s$} &  {\qquad \qquad\qquad$\Delta c$ } \\
			\hline
			$\Xi^{*++}_{cc}(ucc)$ 
			& $1 -a\left(2+ \alpha^2+ \frac{\beta^2}{3}+ \frac{\zeta^2}{24}+ \frac{49}{16}\gamma^2\right)$
			& $- a\left(1+2\gamma^2\right)$
			& $- a \left(\alpha^2+ 2\gamma^2 \right)$
			& $ 2 - \frac{a}{4}\left(3\zeta^2+ 37\gamma^2\right)$ \\\hline

			$\Xi^{*+}_{cc}(dcc)$ 
			& $- a\left(1+2\gamma^2\right)$
			& $1 -a\left(2+ \alpha^2+ \frac{\beta^2}{3}+ \frac{\zeta^2}{24}+ \frac{49}{16}\gamma^2\right)$    
			& $- a \left(\alpha^2+ 2\gamma^2 \right)$
			& $ 2 - \frac{a}{4}\left(3\zeta^2+ 37\gamma^2\right)$ \\\hline

			$\Omega^{*+}_{cc}(ccs)$ 
			& $- a\left(\alpha^2+2\gamma^2\right)$
			& $- a\left(\alpha^2+ 2\gamma^2\right)$
			& $1-a\left(2\alpha^2+ \frac{4}{3} \beta^2+\frac{\zeta^2}{24}+\frac{49}{16}\gamma^2\right)$
			& $2- \frac{a}{4}\left(3\zeta^2+37\gamma^2\right)$ \\\hline
			

			$\Omega^{*++}(ccc)$ 
			&$-3a\gamma^2$ 
			&$-3a\gamma^2$ 
			&$- 3a\gamma^2$ 
			&$3- \frac{9}{8}a\left(\zeta^2+11\gamma^2\right)$
			\\\hline
			
		\end{tabular}
		\caption{\scriptsize The quark spin polarization for doubly  and triply charmed   spin $\frac{3}{2}^{+}$ baryons, (3,2) and (1,3), within $SU(4)$ representation in the chiral quark constituent model.}\label{table_double_charm_3_2}
	\end{center}

\end{sidewaystable}

\begin{table}
	\begin{tabular}{|c|c|c|c|c|c|c|c|c|}
		\hline
		\hline
		Baryon	&	$\Delta u_B$	&	$\Delta d_B$	&	$\Delta s_B$	&	$\Delta c_B$	&	$g_{B}^{0}$	&	$g_{B}^{3}$	&	$g_{B}^{8}$	&	$g_{B}^{15}$	\\
		\hline
		$p$	&	0.939	&	-0.335	&	-0.024	&	-0.001	&0.579	&	1.274	&	0.652	&	0.583	\\
		$n$	&	-0.335	&	0.939	&	-0.024	&	-0.001	&0.579	&	-1.274	&	0.652	&	0.583	\\
		$\Sigma^+$	&	0.911	&	-0.144	&	-0.264	&	-0.001	&0.502	&	1.055	&	1.295	&	0.506	\\
		$\Sigma^0$	&	0.385	&	0.385	&	-0.264	&	-0.001	&0.505	&	0.0	&	1.298	&	0.509	\\
		$\Sigma^-$	&	-0.144	&	0.911	&	-0.264	&	-0.001	&0.502	&	-1.055	&	1.295	&	0.506	\\
		$\Lambda$	&	0.005	&	0.005	&	0.841	&	-0.001	&0.85	&	0.0	&	-1.672	&	0.854	\\
		$\Xi^0$	&	-0.215	&	0.0	&	1.154	&	-0.001	&0.938	&	-0.215	&	-2.523	&	0.942	\\
		$\Xi^-$	&	0.0	&	-0.215	&	1.154	&	-0.001	&0.938	&	0.215	&	-2.523	&	0.942	\\
		\hline
	\end{tabular}
	\caption{ \scriptsize The quark spin polarization and the axial coupling constants for spin $\frac{1}{2}^{+}$ octet baryons (8,0) within  $SU(4)$ representation in the chiral quark constituent model.}  \label{QuarkPolar_Spin_1by2_Octet}
\end{table}

\begin{table}
	\begin{tabular}{|c|c|c|c|c|c|c|c|c|}
		\hline
		\hline
		Baryon	&	$\Delta u_B$	&	$\Delta d_B$	&	$\Delta s_B$	&	$\Delta c_B$	&	$g_{B}^{0}$	&	$g_{B}^{3}$	&	$g_{B}^{8}$	&	$g_{B}^{15}$	\\
		\hline
		$\Sigma_c^{++}$	&	0.908	&	-0.15	&	-0.03	&	-0.249	&0.479	&	1.058	&	0.818	&	1.475	\\
		$\Sigma_c^{+}$	&	0.377	&	0.379	&	-0.03	&	-0.249	&0.477	&	-0.002	&	0.816	&	1.473	\\
		$\Sigma_c^{0}$	&	-0.15	&	0.908	&	-0.03	&	-0.249	&0.479	&	-1.058	&	0.818	&	1.475	\\
		$\Xi_c^{\prime +}$	&	0.439	&	-0.09	&	0.559	&	-0.249	&0.659	&	0.529	&	-0.769	&	1.655	\\
		$\Xi_c^{\prime 0}$	&	-0.09	&	0.439	&	0.559	&	-0.249	&0.659	&	-0.529	&	-0.769	&	1.655	\\
		$\Omega_c^{0}$	&	-0.03	&	-0.03	&	1.149	&	-0.249	&0.84	&	0.0	&	-2.358	&	1.836	\\
		\hline
	\end{tabular}
	\caption{\scriptsize The quark spin polarization and the axial coupling constants for singly charmed spin $\frac{1}{2}^{+}$  baryons (6,1) within  $SU(4)$ representation in the chiral quark constituent model} \label{QuarkPolar_Spin_1by2_Sextet}
\end{table}

\begin{table}
	\begin{tabular}{|c|c|c|c|c|c|c|c|c|}
		\hline
		\hline
		Baryon	&	$\Delta u_B$	&	$\Delta d_B$	&	$\Delta s_B$	&	$\Delta c_B$	&	$g_{B}^{0}$	&	$g_{B}^{3}$	&	$g_{B}^{8}$	&	$g_{B}^{15}$	\\
		\hline
		$\Lambda_c^{+}$	&	0.02	&	0.02	&	-0.003	&	0.893	&0.93	&	0.0	&	0.046	&	-2.642	\\
		$\Xi_c^+$	&	0.026	&	-0.007	&	0.033	&	0.893	&0.945	&	0.033	&	-0.047	&	-2.627	\\
		$\Xi_c^0$	&	-0.007	&	0.026	&	0.033	&	0.893	&0.945	&	-0.033	&	-0.047	&	-2.627	\\
		$\Xi_{cc}^{++}$	&	-0.187	&	0.029	&	0.004	&	1.216	&1.062	&	-0.216	&	-0.166	&	-3.802	\\
		$\Xi_{cc}^{+}$	&	0.029	&	-0.187	&	0.004	&	1.216	&1.062	&	0.216	&	-0.166	&	-3.802	\\
		$\Omega_{cc}^+$	&	0.004	&	0.004	&	-0.235	&	1.216	&0.989	&	0.0	&	0.478	&	-3.875	\\
		\hline
	\end{tabular}
	\caption{\scriptsize The quark spin polarization and the axial coupling constants for double charmed spin $\frac{1}{2}^{+}$ baryons (3,2) within $SU(4)$ representation in the chiral quark constituent model.} \label{QuarkPolar_Spin_1by2_triplet}
\end{table}

\begin{table}
	
	\begin{tabular}{|c|c|c|c|c|c|c|c|c|}
		\hline
		\hline
		Baryon        &       $\Delta u_B$    &       $\Delta d_B$    &       $\Delta s_B$    &       $\Delta c_B$    &       $g_{B}^{0}$   &       $g_{B}^{3}$   &       $g_{B}^{8}$   &       $g_{B}^{15}$  \\
		\hline
		$\Delta^{++}$   &       2.17    &       -0.36   &       -0.073  &       -0.004  &1.733  &       2.53    &       1.956   &       1.749   \\
		$\Delta^{+}$    &       1.326   &       0.483   &       -0.073  &       -0.004  &1.732  &       0.843   &       1.955   &       1.748   \\
		$\Delta^{0}$    &       0.483   &       1.326   &       -0.073  &       -0.004  &1.732  &       -0.843  &       1.955   &       1.748   \\
		$\Delta^{-}$    &       -0.36   &       2.17    &       -0.073  &       -0.004  &1.733  &       -2.53   &       1.956   &       1.749   \\
		$\Sigma^{\star +}$      &       1.422   &       -0.264  &       0.866   &       -0.004  &2.02   &       1.686   &       -0.574  &       2.036   \\
		$\Sigma^{\star 0}$      &       0.579   &       0.579   &       0.866   &       -0.004  &2.02   &       0.0     &       -0.574  &       2.036   \\
		$\Sigma^{\star -}$      &       -0.264  &       1.422   &       0.866   &       -0.004  &2.02   &       -1.686  &       -0.574  &       2.036   \\
		$\Xi^{\star 0}$ &       0.675   &       -0.169  &       1.805   &       -0.004  &2.307  &       0.844   &       -3.104  &       2.323   \\
		$\Xi^{\star -}$ &       -0.169  &       0.675   &       1.805   &       -0.004  &2.307  &       -0.844  &       -3.104  &       2.323   \\
		$\Omega^{-}$    &       -0.073  &       -0.073  &       2.744   &       -0.004  &2.594  &       0.0     &       -5.634  &       2.61    \\

		\hline
	\end{tabular}                                                                                                                              
	\caption{The quark spin polarization and the axial coupling constants for spin $\frac{3}{2}^{+}$,  decuplet baryons (10,0) within SU (4) representation in the chiral quark constituent model.}

	\label{QuarkPolar_Spin_3by2_decuplet}
\end{table}

\begin{table}[h]
	\begin{tabular}{|c|c|c|c|c|c|c|c|c|}
		\hline
		\hline
		Baryon        &       $\Delta u_B$    &       $\Delta d_B$    &       $\Delta s_B$    &       $\Delta c_B$    &       $g_{B}^{0}$   &       $g_{B}^{3}$   &       $g_{B}^{8}$   &       $g_{B}^{15}$  \\
		\hline
		$\Sigma_c^{\star ++}$   &       1.445   &       -0.241  &       -0.05   &       0.966   &2.12   &       1.686   &       1.304   &       -1.744  \\
		$\Sigma_{c}^{\star +}$  &       0.602   &       0.602   &       -0.05   &       0.966   &2.12   &       0.0     &       1.304   &       -1.744  \\
		$\Sigma_{c}^{\star 0}$  &       -0.241  &       1.445   &       -0.05   &       0.966   &2.12   &       -1.686  &       1.304   &       -1.744  \\
		$\Xi_{c}^{\star +}$     &       0.697   &       -0.146  &       0.889   &       0.966   &2.406  &       0.843   &       -1.227  &       -1.458  \\
		$\Xi_{c}^{\star 0}$     &       -0.146  &       0.697   &       0.889   &       0.966   &2.406  &       -0.843  &       -1.227  &       -1.458  \\
		$\Omega_{c}^{\star 0}$  &       -0.05   &       -0.05   &       1.828   &       0.966   &2.694  &       0.0     &       -3.756  &       -1.17   \\
		
		\hline
	\end{tabular}                                                                                                                              
	\caption{The quark spin polarization and the axial coupling constants for spin $\frac{3}{2}^{+}$,  sextet baryons (6,1) within SU (4) representation in the chiral quark constituent model.}
	
	\label{QuarkPolar_Spin_3by2_Sextet}
\end{table}

\begin{table}[h]
	
	\begin{tabular}{|c|c|c|c|c|c|c|c|c|}
		\hline
		\hline
		Baryon        &       $\Delta u_B$    &       $\Delta d_B$    &       $\Delta s_B$    &       $\Delta c_B$    &       $g_{B}^{0}$   &       $g_{B}^{3}$   &       $g_{B}^{8}$   &       $g_{B}^{15}$  \\
		\hline
		$\Xi_{cc}^{\star ++}$   &       0.72    &       -0.123  &       -0.027  &       1.936   &2.506  &       0.843   &       0.651   &       -5.238  \\
		$\Xi_{cc}^{\star +}$    &       -0.28   &       0.877   &       -0.027  &       1.936   &2.506  &       -1.157  &       0.651   &       -5.238  \\
		$\Omega_{cc}^{\star +}$ &       -0.027  &       -0.027  &       0.912   &       1.936   &2.794  &       0.0     &       -1.878  &       -4.95   \\
		$\Omega_{ccc}^{\star ++}$       &       -0.004  &       -0.004  &       -0.004  &       2.906   &2.894  &       0.0     &       0.0     &       -8.73   \\
		
		\hline
	\end{tabular}                                                                                                                              
	\caption{The quark spin polarization and the axial coupling constants for spin $\frac{3}{2}^{+}$,  triplet baryons (3,2) and singlet (1,3) baryons within SU (4) representation in the chiral quark constituent model.}
	
	\label{QuarkPolar_Spin_3by2_triplet_and_singlet}
\end{table}


\end{document}